\documentclass[11pt]{article}  
\usepackage{graphicx}

\usepackage{amsmath,amsthm}
\usepackage{amssymb}
\usepackage{amsfonts}

\usepackage{color}
\input epsf
\parskip=\medskipamount
   \def\CaH{{\cal H}}
  \def\CaK{{\cal K}} \def\CaL{{\cal L}}

\def\al{\alpha}
\def\be{\beta}

\def\la{\lambda}  
  
\def\kp{\lambda}

\def\IB{\relax{\rm l\kern-.18 em B}}
\def\IC{\relax{\rm l\kern-.50 em C}}
\def\IE{\relax{\rm l\kern-.12 em E}}
\def\IH{\relax{\rm l\kern-.18 em H}}
\def\IK{\relax{\rm l\kern-.18 em K}}
\def\IL{\relax{\rm I\kern-.18 em L}}
\def\IN{\relax{\rm I\kern-.18 em N}}
\def\IR{\relax{\rm I\kern-.18 em R}}

\def\\{\hfill\break}

\def\wt{\widetilde}

\font\tenfrak=eufm10  \font\sevenfrak=eufm7  \font\fivefrak=eufm5
\newfam\frakfam
\textfont\frakfam=\tenfrak
\scriptfont\frakfam=\sevenfrak
\scriptscriptfont\frakfam=\fivefrak

\newtheorem{proposition}{Proposition}

\def\wt{\widetilde}
\def\fracpd#1#2{\frac{\partial #1}{\partial #2}}
\def\ptos{\leaders\hbox to 2mm{\hfil{.}\hfil}\hfill}

\newcommand{\sect}[1]{\setcounter{equation}{0}\section{#1}}
\newcommand{\subsect}[1]{\subsection{#1}}
\newcommand{\subsubsect}[1]{\subsubsection{#1}}
\renewcommand{\theequation}{\arabic{section}.\arabic{equation}}

\def\dd{{\rm d}}
\def\pot{{\cal U}}

\def\pa{\tau}
\def\pb{\sigma}

\def\beq{\begin{equation}}
\def\eeq{\end{equation}}
\def\bea{\begin{eqnarray}}
\def\eea{\end{eqnarray}}

\begin{document}

\title{Superintegrable  systems on 3-dimensional curved spaces: \\ 
Eisenhart formalism and separability }

\author{ Jose F. Cari\~nena$^{\dagger\,a)}$,
Francisco J. Herranz$^{\ddagger\,b)}$,
Manuel F. Ra\~nada$^{\dagger\,c)}$  \\ [3pt]
${}^{\dagger}$
   {\it Departamento de F\'{\i}sica Te\'orica and IUMA, Facultad de Ciencias} \\
   {\it Universidad de Zaragoza, 50009 Zaragoza, Spain}  \\   
${}^{\ddagger}$
   {\it Departamento de F\'{\i}sica, Universidad de Burgos, 09001 Burgos, Spain}  }
\maketitle

\begin{abstract} 
The Eisenhart geometric formalism, which  transforms  an Euclidean natural Hamiltonian $H=T+V$  into a geodesic Hamiltonian ${\cal T}$  with one additional degree of freedom, is applied to the four families of quadratically superintegrable systems with multiple separability in the Euclidean plane. 
Firstly,  the separability and superintegrability of such four   geodesic Hamiltonians  ${\cal  T}_r$  ($r=a,b,c,d$) in a three-dimensional curved space are studied and then these four systems are modified  with the addition of a potential  ${\cal U}_r$ leading to ${\cal H}_r={\cal  T}_r +{\cal U}_r$. 
Secondly, we  study  the superintegrability of the four Hamiltonians $\wt{{\cal H}}_r= {\cal H}_r/ \mu_r$, where   $\mu_r$ is a certain position-dependent mass, that enjoys the same  separability as the original system ${\cal H}_r$. 
All the Hamiltonians  here studied  describe superintegrable systems on non-Euclidean  three-dimensional manifolds  with a broken spherically symmetry.
\end{abstract}

\begin{quote}
Position dependent mass. Geodesic Hamiltonians. Constants of motion. Laplace--Runge--Lenz vector.  
 

\end{quote}

\vfill
\footnoterule
{\noindent\small
$^{a)}${\it E-mail address:} {\tt  jfc@unizar.es } \\
$^{b)}${\it E-mail address:} {\tt  fjherranz@ubu.es }  \\
$^{c)}${\it E-mail address:} {\tt  mfran@unizar.es }
}
\newpage


\sect{Introduction}

 It is well known that the harmonic (isotropic)  oscillator and the Kepler--Coulomb (KC)  problem are integrable systems admitting  additional constants of motion (Demkov--Fradkin tensor~\cite{Demkov,  Fradkin} and Laplace--Runge--Lenz vector, respectively).
Systems endowed with this property are called superintegrable.  
It is also known that if a system is separable (Hamilton--Jacobi (HJ) separable in the classical case or Schr\"odinger separable in the quantum case), then it  is integrable with integrals of motion of at most second-order in momenta.   
Thus, if a system admits multiseparability (separability in several different systems of coordinates) then it is endowed with `quadratic superintegrability'  (superintegrability with linear or quadratic integrals of motion). 

 Fris  {\it et al}.~studied in  \cite{FrMaSmUW65}  the two-dimensional  (2D) Euclidean systems  admitting   separability in more than one coordinate system and they obtained four families of potentials $V_r$, $r = a, b, c, d$, possessing three functionally independent integrals of motion  (they were mainly interested in the quantum 2D Schr\"odinger equation but their results     also hold at the classical level). 
 Then other authors studied similar problems on higher-dimensional Euclidean spaces \cite{Ev90Pra}--\cite{KaWiMiPo99}, on 2D spaces with a pseudo-Euclidean metric (Drach potentials) \cite{Ra97Jmp}--\cite{Camp14}, and  on curved spaces \cite{GrPoSi95b}--\cite{Mfran15PLa} (see \cite{MiPWJPa13} for a recent review on superintegrability that includes a long list of references). 
 
  The  superintegrability property is related with different formalisms and it can be studied by making use of different approaches, that is,  proving that all bounded classical trajectories are closed, HJ  separability, action-angle variables formalism, exact solvability, degenerate quantum  energy levels, complex functions whose Poisson bracket with the Hamiltonian are proportional to themselves, etc. In this paper, we relate superintegrability with a geometric formalism introduced  many years ago by Eisenhart  \cite{Eisenhart28}.

The  theory of  general  relativity states that the motion of a particle under the action of gravitational forces is described by a geodesic in the 4D  Riemannian spacetime. 
The Eisenhart formalism (also known as Eisenhart lift) associates to a system governed by a natural   Hamiltonian   $H=T+V$ (a kinetic term plus a potential) a new  geodesic  Hamiltonian $\cal T$  (so without any potential) with an additional degree of freedom (it is in fact an extended formalism). 
The important point is that  the solutions  of the equations of motion for such a Hamiltonian $H$ come from   geodesics of  $\cal T$ in an enlarged curved space. 
That is, it is a geometric formalism introduced with the idea of relating classical nonrelativistic  Lagrangian or Hamiltonian mechanics with relativistic gravitation \cite{Szydlowski98}--\cite{CarigliaAlv15}. 
Our idea is that this formalism can also  be applied for the study of superintegrable systems on non-Euclidean  spaces.

  One important point is that although the number of superintegrable systems can be considered as rather limited, they are not,  however, isolated ones but, on the contrary, they frequently appear grouped  into  families; for example, each of the above  mentioned 2D  potentials $V_r$ ($r=a,b,c,d$), has  the structure of a  3D vector space. 
In this paper we prove  that the 2D Euclidean potentials $V_r$ are related, via the Eisenhart formalism, with some superintegrable   geodesic Hamiltonian  systems ${\cal T}_r$  on 3D curved spaces, generally of nonconstant curvature and with a broken spherically symmetry. Furthermore, natural 3D Hamiltonians,  ${\cal H}_r={\cal T}_r+{\cal U}_r$, can then be constructed by preserving the same  superintegrability and separability properties.

  On the other hand,  in these last years the interest for the study of systems with a  position-dependent mass (PDM) has become a matter of great interest and has attracted a lot of attention of many authors 
\cite{Vak05}--\cite{Quesne15Jmp}.  
It seems therefore natural to enlarge the study of superintegrability and separability to include systems with a PDM by following the same constructive approach. 
Consequently, as a new step in this procedure, we also  prove  that ${\cal T}_r$  and  ${\cal H}_r$ admit    deformations, say  $\wt{\cal T}_r={\cal T}_r/\mu_r$  and  $\wt{\cal H}_r={\cal H}_r/\mu_r$, with a  PDM  $\mu_r(\kp)$ depending of a real parameter $\kp$,  in such a way that the latter are superintegrable for all the values of $\kp$ (in the domain of the parameter) and that for $\kp=0$ they reduce to the 
  previously studied superintegrable  Hamiltonians.

We must mention that there exists  a certain relationship between the approach presented in this paper and some previous studies on  curved  oscillators and KC potentials related to the so-called Bertrand spacetimes 
(spherically symmetric and static Lorentzian spacetimes), firstly introduced  by Perlick in \cite{Perlick92} and further studied in~\cite{Ballestetal08ClQGr, RagRig10}, where  generalisations of superintegrable Hamiltonians  fulfilling     Bertrand's theorem~\cite{ComRunge} on  conformally flat  spaces have been achieved. 
We stress that  one of the main differences (in addition to the use of the Eisenhart formalism) is that,   in this paper the potentials are not necessary central, that is,  our results mainly concern  systems defined on non-conformally flat   spaces.

  The structure of the paper is as follows.  In the next section we establish  the main characteristics of the Eisenhart formalism (a rigorous geometrical description can be found in the Appendix). 
   In Section 3 we briefly review the classification of the four families of   quadratic in the momenta  superintegrable Hamiltonians on the Euclidean plane $H_r=T+V_r$ $(r=a,b,c,d)$. In Section 4, the Eisenhart approach is applied in order to construct superintegrable  3D geodesic Hamiltonians ${\cal T}_r$ from the previous $V_r$. 
   The addition of a potential $\pot_r$ to ${\cal T}_r$ leading to    superintegrable/separable  Hamiltonians  ${\cal H}_r={\cal T}_r+\pot_r$ is addressed in Section 5. Next in Section 6, a  PDM $\mu_r$ is   introduced in    the 3D geodesic Hamiltonians   ${\cal T}_r$,  by preserving     separability, so  giving rise to   new superintegrable geodesic Hamiltonians $\wt{\cal T}_r={\cal T}_r/\mu_r$. In Section 7, a  separable potential $\wt\pot_r$ is added to  $\wt{\cal T}_r$ providing new  superintegrable Hamiltonians  $\wt{\cal H}_r$ which constitute  the main result of this paper.  
We conclude in the last section with some remarks and open problems.

\sect{Eisenhart formalism }

Let us first recall some basic properties relating Riemannian geometry with Lagrangian dynamics for natural systems.

Suppose a $n$D manifold $M$ endowed with a Riemannian metric $g$. If we denote by $\{q^i \,; i=1,\ldots,n\}$, a set of coordinates on $M$ and by  $g_{ij}(q)$ the components of $g$,   the expressions of $g$ and $\dd s^2$ are given by
$$
  g = g_{ij}(q)\,\dd q^i \otimes \dd q^j  ,\qquad   \dd s^2  = g_{ij}(q) \dd q^i \dd q^j  \,. 
$$
Then the corresponding equation of the geodesics on $M$, 
$$
\ddot q^i+\Gamma^i_{jk}\,\dot q^j\,\dot q^k=0  \,,{\qquad} 
\Gamma^i_{jk}=\frac 12 \, g^{il} \left(\fracpd{g_{lj}}{q^k}+\fracpd{g_{lk}}{q^j}-\fracpd{g_{jk}}{q^l}\right) ,\quad i,j,k= 1,\ldots,n,
$$
can be obtained as the Euler--Lagrange equations from a Lagrangian $L$ with only a quadratic kinetic term $T_g$ and without any potential 
$$
   {L = T_g }=  \frac 12\,g_{ij}(q) v^i v^j  \,. 
$$
Conversely,  the Lagrangian formalism establishes that the trajectories of free motion of a particle in a configuration space $Q$ are (i) the solutions of the equations determined by a pure kinetic Lagrangian (quadratic kinetic term without potential) and that (ii) these trajectories are just the geodesics on the space $Q$. 
Hence the Lagrangians describing the free motion are also known as geodesic Lagrangians.

 As it is well known, the relativistic theory of gravitation introduced by Einstein in 1915 establishes that the trajectory of a particle  under external gravitational forces can be described as a geodesic on the 4D  spacetime. 
 This, in turn, means that the spatial paths of particles in the 3D Euclidean space can alternatively be considered as geodesics in a higher-dimensional non-Euclidean space  by introducing a new metric. 
This  was the idea introduced later on  by Eisenhart  in 1928/29 in nonrelativistic Lagrangian and Hamiltonian dynamics~\cite{Eisenhart28}. Thus   the equations of motion of a particle under the action of a potential force  in a $n$D configuration space $Q$ can be reformulated as the equations of geodesics in a $(n+1)$D new configuration space $\wt{Q}$ with a new (pseudo-)Riemannian metric constructed by combining  the original metric with the potential defined on $Q$. 
 
More explicitly, assume that we are given a natural Lagrangian (quadratic kinetic term minus a potential) 
$$
  L (q,v)= T (q,v) - V(q) \,,{\qquad} T (q,v)=  \frac 12\,g_{ij}(q) v^i v^j  \, ,
$$
where the coefficients $g_{ij}(q)$ are symmetric functions of the coordinates and $V(q)$ is a potential. We can then consider the configuration space $Q$ of the system as a Riemannian space with a metric  determined by the coefficients of the kinetic term
$$
 \dd s^2  = g_{ij}(q) \dd q^i \dd q^j  \, .
 $$
Since the matrix $[g_{ij}(q)]$ is invertible the Legendre transformation, $p_i= g_{ij}(q) \,v^j$, leads to the Hamiltonian function $H$ given by 
$$
  H (q,p)= T(q,p) + V(q) \,,{\qquad} T =  \frac 12\,g^{ij}(q) p_i p_j  \,,
$$
where $ g_{ij}\, g^{jk}=\delta_i^k$.

The Eisenhart formalism (also known as Eisenhart lift) is an extended formalism. 
The main idea is introducing a new degree of freedom with a new coordinate, say $z$, i.e. $Q$ is replaced by $\wt{Q}=\mathbb{R}\times Q$  and its corresponding momentum $p_z$, 
in such a way that the new metric $\dd\sigma^2$ and the new   Hamiltonian $ {\cal T}\in C^\infty(T^*{\wt{Q}}) $ are given by
\beq
 \dd\sigma^2  = g_{ij}(q) \dd q^i \dd q^j + \frac{\dd z^2}{V(q)}  
 \,,{\qquad}
  {\cal T} = \frac 12\,g^{ij}(q) p_i p_j + \frac 12\, V(q) \, p_z^2  \, ,
\label{ba}
\eeq
so that ${\cal T}$ is homogeneous of  degree  two in the momenta, and this defines  a geodesic Hamiltonian.
As the variable $z$ is cyclic $p_z$ is a constant of motion  and fixing its value $p_z=1$ the parameter of the integral curves coincides with the arc-length. 
In this way the motion of a particle under external forces arising from a potential $V$ is described as  a geodesic motion in an extended configuration space determined by  ${\cal T}$. 
Although the origin of this formalism is related with properties of relativistic mechanics, this procedure has been studied  by  making use of different approaches (see \cite{Szydlowski98}--\cite{CarigliaAlv15} and references therein). 
 A more detailed geometric study of the Eisenhart lift is presented in the Appendix. 

  We must mention that Eisenhart also considered another more extended formalism  that introduces not just one but two additional degrees of freedom; that is, two new variables ($Q$ is replaced by $\wt{Q} =\mathbb{R}^2\times Q$) and two conjugated momenta \cite{Eisenhart28}, \cite{Cariglia14Rmp}. 
 This more genera Eisenhart lift is related with the study of time-dependent systems and with problems with external gauge fields (Lagrangians with terms linear in the velocities).
 Nevertheless in what follows we study time-independent natural Hamiltonians without gauge fields and, therefore (see Section 6 of [23]),  we will make use the Eisenhart formalism with only one extra degree of freedom.

\sect{Quadratic superintegrability in the Euclidean plane}

Let us denote by   $V_r$, $r=a,b,c,d$,  the four 2D potentials with separability in two different coordinate systems in the Euclidean plane  \cite{FrMaSmUW65}--\cite{KaWiMiPo99}.  
Each resulting potential $V_r$ is, in fact,  a superposition of three potentials
$$
V_r=k_1 V_1+ k_2 V_2 +k_3 V_3 \, ,
$$
where, hereafter, $k_1,k_2,k_3$  are three  arbitrary real constants. We remark that, from a mathematical/physical viewpoint,  the $k_1$-term will be the `principal' potential   so that each family will be `shortly' named according to it.

For our purposes we write these four families in terms of Cartesian coordinates $(x,y)$ with conjugate momenta $(p_x,p_y)$, in such a manner that the Hamiltonian $H_r$ reads
\beq
H_r=T+V_r=  \frac 12  \, (p_x^2+p_y^2)+V_r(x,y)\, , \qquad r=a,b,c,d\, .
\label{ca}
\eeq
These four  types of Hamiltonians determine quadratically  superintegrable systems as  they are endowed with {\em  three} functionally independent constants of motion which are quadratic in the momenta. 
Notice that for $n=2$, the superintegrability property is, in fact,  maximal since $2n-1=3$ is the maximum number of independent integrals.

   The two first potentials, $V_a$ and $V_b$, represent nonlinear oscillators (harmonic oscillators  with additional terms), meanwhile the  two remaining potentials, $V_c$ and $V_d$, correspond to the superposition of  the KC   problem with two other terms.

\subsect{Family a: Isotropic  oscillator}  
This corresponds to the potential  
\begin{equation}   
  V_a =  \frac 12\,k_1 (x^2 + y^2) + \frac{k_2}{x^2}  + \frac{k_3}{y^2}\,  ,
  \label{cb}
\end{equation}
 which   is separable in  (i) Cartesian coordinates and  (ii) polar ones. 
 The $k_1$-potential is just the isotropic oscillator with frequency $\omega$ whenever $k_1=\omega^2>0$, meanwhile the two remaining potentials are   Rosochatius  or Winternitz  terms (which provide centrifugal barriers when $k_2>0$ and $k_3>0$). 
  We recall that  the Hamiltonian  $H_a$ is just the 2D version  of the so-called Smorodinsky--Winternitz system~\cite{FrMaSmUW65} which has been widely studied  (see, e.g.,~\cite{GrPoSi95a, KaWiMiPo99, BaHeSantS03, Evans90, Evans91, LetterBH} and references therein).

Three functionally independent  constants of motion are the two 1D energies, $I_{a1}$ and $I_{a2}$, along with a third integral $I_{a3}$ related to the angular momentum; namely, 
\bea
&& I_{a1}= {\frac 12 }\,  p_x^2 + {\frac 12 }\,k_1 x^2  + \frac{k_2}{x^2}  \, ,\qquad I_{a2}=  {\frac 12 }\,  p_y^2 +  {\frac 12 }\,k_1 y^2  + \frac{k_3}{y^2}  \, ,
\nonumber\\[2pt]
&&  I_{a3}  =   (x p_y - y p_x)^2  +  2k_2\biggl(\frac{y}{x}\biggr)^2 +
  2k_3\left(\frac{x}{y}\right)^2   .
\nonumber
\eea

\subsect{Family b: Anisotropic oscillator}   
The following potential  
\begin{equation}  
  V_b = \frac 12 \, k_1 (4 x^2 + y^2) + \frac{k_2}{y^2}  + k_3 x
  \label{cc}
\end{equation}
is separable in  (i) Cartesian coordinates and  (ii) parabolic ones. The $k_1$-potential is just the anisotropic $2:1$ oscillator provided that $k_1=\omega^2>0$ (so with frequencies $\omega_x=2\omega$ and $\omega_y=\omega$), the $k_2$-potential is a   Rosochatius--Winternitz term, and the (trivial)  $k_3$-potential simply  corresponds to a translation along the $x$-axis.

Three constants of motion are the two 1D energies, $I_{b1}$, $I_{b2}$, and a third integral $I_{b3}$,  related to one component of the 2D Laplace--Runge--Lenz vector, which are given by 
\bea
&&I_{b1} =   {\frac 12 }\,  p_x^2 + 2  k_1   x^2    + k_3 x \, , \qquad I_{b2} =  {\frac 12 }\,  p_y^2 + \frac 12\,  k_1  y^2 + \frac{k_2}{y^2}   \, , \nonumber\\[2pt]
&&  I_{b3} =  (x p_y - y p_x) p_y  -  k_1 x y^2 +  \frac{ 2 k_2   x}{y^2}  - \frac{k_3 y^2}{2}
  \,. 
  \nonumber
\eea

\subsect{Family c: Kepler--Coulomb I} 
The   potential  given by
\begin{equation}   
  V_c =  \frac{k_1}{\sqrt{x^2 + y^2}}  +  \frac{k_2}{y^2}  +  \frac{k_3x}{y^2 \sqrt{x^2 + y^2}}
  \label{cd}
\end{equation}   
is separable in  (i) polar coordinates and  (ii) parabolic ones. In this case, the $k_1$-term is the KC potential and the $k_2$-term is a  Rosochatius--Winternitz potential.

One constant of motion is the Hamiltonian itself, that is $ I_{c1} =  H_c$, and   two other integrals,  $I_{c2}$, and $I_{c3}$, read   
\begin{eqnarray}
&&  I_{c2}  =  (x p_y - y p_x)^2 + \frac{2 k_2 x^2}{y^2} +
  \frac{2 k_3 x \sqrt{x^2 + y^2}}{y^2}  \,, \cr
&&  I_{c3}= (x p_y - y p_x) p_y + \frac{k_1 x}{\sqrt{x^2 + y^2}}
  + \frac{2 k_2 x}{y^2} + \frac{k_3 (2 x^2 + y^2)}{y^2 \sqrt{x^2 +
y^2}}   \,. \nonumber
\end{eqnarray}
Hence $ I_{c2}$ comes from the angular momentum, while  $ I_{c3}$ is provided by a component of the   Laplace--Runge--Lenz vector.

\subsect{Family d: Kepler--Coulomb II}
Finally,  the fourth potential  is given by
\begin{equation}   
V_d =  \frac{k_1}{\sqrt{x^2 + y^2}}   
  + k_2 \, \frac{ \bigl[\sqrt{x^2 + y^2} + x\bigr]^{1/2} }{ \sqrt{x^2 + y^2}}   
  + k_3 \,\frac{ \bigl[\sqrt{x^2 + y^2} - x\bigr]^{1/2} }{ \sqrt{x^2 + y^2}} \, ,
  \label{ce}
\end{equation}   
 which is separable in  (i) parabolic coordinates $(\pa,\pb)$ and  (ii) a second system of parabolic coordinates $(\alpha,\beta)$ obtained from $(\pa,\pb)$ by a rotation. 
 Thus, we recall that  the KC potential ($k_1$-term)  can be superposed with two other  potentials which are different from the previous ones   (\ref{cd})  keeping superintegrability.

One constant of motion is  again the Hamiltonian itself,  $ I_{d1} =  H_d$, meanwhile   two other integrals,  $I_{d2}$, and $I_{d3}$, turn out to be
 \begin{eqnarray}
  I_{d2} \! \! \!&=&\! \!\!  (x p_y - y p_x) p_y + \frac{k_1 x}{\sqrt{x^2 + y^2}}
-   \frac{k_2 y\, \bigl[\sqrt{x^2 + y^2} - x\bigr]^{1/2} }{\sqrt{x^2 + y^2}} 
+  \frac{k_3 y\, \bigl[\sqrt{x^2 + y^2} + x\bigr]^{1/2} }{\sqrt{x^2 + y^2}}  \,,   \nonumber  \\[2pt]
  I_{d3}  \! \! \!&=&\! \!\!  (x p_y - y p_x) p_x - \frac{k_1 y}{\sqrt{x^2 + y^2}}
-  \frac{ k_2 x\, \bigl[\sqrt{x^2 + y^2} - x\bigr]^{1/2} } {\sqrt{x^2 + y^2}}  
+  \frac{ k_3 x\, \bigl[\sqrt{x^2 + y^2} + x\bigr]^{1/2} } {\sqrt{x^2 + y^2}} \,.   \nonumber  
\end{eqnarray}
These are related to both components of the  2D Laplace--Runge--Lenz vector.

Obviously, when $k_2=k_3=0$ both KC I and II families  reduce to the common  KC $k_1$-potential.  Nevertheless, throughout the paper we shall deal with the three generic $k_i$-terms so describing two essential different families of superintegrable Hamiltonians.

\sect{Geodesic Hamiltonians ${\cal T}$ endowed with multiple separability on 3D curved spaces }

Let  us consider  the  2D Euclidean Hamiltonian $H_r$  (\ref{ca})   with one of the superintegrable potentials $V_r$ given in the above section.  By applying the Eisenhart lift (\ref{ba})  with $g_{ij}=\delta_{ij}$,  $q_1=x, q_2=y$,  we obtain a  new 3D Riemannian metric and associated   free Hamiltonian   defined  by  
\begin{equation}   
 \dd \sigma_r^2=   \dd x^2 +  \dd y^2 + \frac{\dd z^2}{V_r(x,y)} \,  ,\qquad 
 {\cal T}_r = \frac{1}{2}\left( p_x^2 + p_y^2 + V_r(x,y)\,p_z^2 \right) \, , \qquad r=a,b,c,d\, ,  \label{CaHa}
\end{equation} 
where $(x,y,z)$ are Cartesian coordinates and $(p_x,p_y,p_z)$ their conjugate momenta.

In what follows we study the separability of the corresponding HJ  equation for each of the four types of geodesic Hamiltonians  $ {\cal T}_r $. We stress that the separability of $ {\cal T}_r $ is, in fact,    provided by the separability of $H_r$, that is, if  $H_r$ is separable in the coordinates $(q_1,q_2)$, we assume that $ {\cal T}_r $ is separable in the coordinates $(q_1,q_2,z)$.
We advance that we shall obtain {\em four} independent integrals for each   $ {\cal T}_r $  in an explicit form (three of them being mutually in involution). Consequently,  $ {\cal T}_r $ will determine a superintegrable system but not a maximal superintegrable  one, since an additional {\em fifth} constant of motion would be necessary to get the maximum number of $2n-1=5$ integrals (corresponding to $n=3$   degrees of freedom). 
In this sense,  $ {\cal T}_r $ can be regarded  as either a minimally superintegrable Hamiltonian~\cite{Ev90Pra} or a quasi-maximally superintegrable one~\cite{LetterBH}.

At this point we mention  that the idea of obtaining  a new superintegrable $(n+1)$D Hamiltonian starting with a simpler and  previously known superintegrable Hamiltonian with $n$ degrees of freedom  is a matter that has been  analyzed by some authors (see e.g. \cite{ChanuDgR11,ChanuDgR14,ChanuDgR15}) but making use of other approaches different to the Eisenhart formalism presented in this paper.

\subsect{Geodesic Hamiltonian ${\cal T}_a$ from   isotropic oscillator}

We construct the Hamiltonian  ${\cal T}_a$  (\ref{CaHa})   with the potential $V_a$ (\ref{cb}).   
Since the initial 2D Hamiltonian  $H_a$ is separable in   Cartesian $(x,y)$ and polar variables $(r,\phi)$, we now analyse the separability of the new 3D  system  $ {\cal T}_a $  in  Cartesian  $(x,y,z)$ and cylindrical $(r,\phi,z)$ coordinates.

\subsubsect{Cartesian separability}

 The HJ  equation takes the form 
$$
\left(\fracpd{W}{x}\right)^2 + \left(\fracpd{W}{y}\right)^2  
 +  V_a(x,y) \left(\fracpd{W}{z}\right)^2  = 2E \, ,
$$
so that if we assume that $W$ can be written as $W = W_x(x) + W_y(y) + W_z(z)$, then we can perform a separation of variables which leads  to the following one-variable expressions 
$$
 (W_z')^2   = -\,\alpha  \,,{\quad\ } 
 (W_x')^2  - \alpha\left({\frac 12}\, k_1 x^2 + \frac{k_2}{x^2}\right)  =  \beta +  E   
 \,,{\quad\ } 
  (W_y')^2  - \alpha\left({\frac 12}\,k_1 y^2 + \frac{k_3}{y^2}\right) =   -\,\beta +  E  
  \,,
$$
where $\alpha$ and $\beta$ denote  two constants associated with separability. 
Each one of these expressions determines a constant of motion; so the following functions 
\beq
 K_{a1}  = p_z  \,,{\qquad} 
 K_{a2}  = p_x^2 + \left({\frac 12}\, k_1x^2+\frac{k_2}{x^2}\right)p_z^2     \,,{\qquad} 
 K_{a3} = p_y^2 + \left( {\frac 12}\, k_1y^2+\frac{k_3}{y^2}\right)p_z^2     \,,  
\label{da}
\eeq
are three functionally independent constants of motion,  
$$
   \dd K_{a1}\,\wedge\, \dd K_{a2}\,\wedge\, \dd K_{a3}\ne 0 \,,{\quad} 
$$
satisfying the following properties 
$$
  \{K_{a1}, K_{a2}\} = 0 \,,{\qquad} 
  \{K_{a1}, K_{a3}\} = 0 \,,{\qquad} \{K_{a2},K_{a3}\} = 0 \,,{\qquad} 
  {\cal T}_a  = \frac{1}{2}\bigl(K_{a2} +  K_{a3} \bigr) \,. 
$$

\subsubsect{Cylindrical separability}

We introduce the usual polar coordinates,    $x=r\cos\phi$ and $y=r\sin\phi$,  finding that the free Hamiltonian  $T_a$ reads
$$
 {\cal T}_a = \frac{1}{2}\left( p_r^2 + \frac{p_\phi^2}{r^2} + V_a\,p_z^2 \right) \, , \qquad    V_a =  \frac 12\,k_1 r^2 + \frac{k_2}{r^2\cos^2\phi}  + \frac{k_3}{r^2\sin^2\phi}\,  .
 $$
  The HJ equation turns out to be
$$
  \left(\fracpd{W}{r}\right)^2 + \frac{1}{r^2}\left(\fracpd{W}{\phi}\right)^2 
  +  V_a(r,\phi) \left(\fracpd{W}{z}\right)^2   =  2 E  \,. 
$$
If we suppose that $W = W_r(r) + W_\phi(\phi) + W_z(z)$, then we can perform a separation of variables;  we  first  obtain $(W_z')^2   = -\,\gamma$ and next 
$$
   r^2(W_r')^2   - 2 r^2 E  -  \frac 12\,k_1r^4 \gamma = 
   -\, (W_\phi')^2 + \gamma  \left(   \frac{k_2}{\cos^2\phi}  +  \frac{k_3}{\sin^2\phi}\right) = \delta ,
$$
where $\gamma$ and $\delta$ are two constants. Hence we obtain the following constants of motion  
\beq
J_{a1}= p_z\equiv  K_{a1}   \,  , \quad\  J_{a2} = p_\phi^2  +  \left(   \frac{k_2}{\cos^2\phi}  +  \frac{k_3}{\sin^2\phi}\right) p_z^2\,, \quad\ 
 J_{a3}= r^2 p_r^2 +\frac 12\,k_1r^4 p_z^2   -  2 r^2 {\cal T}_a\,,  
 \label{db}
\eeq
such that  $\{J_{a1}\,,J_{a2}\} = 0$  and  $J_{a2}+  J_{a3} = 0 $.

We summarize the above results in the following statement. 
\begin{proposition}  \label{proposition1}
The 3D geodesic Hamiltonian 
\beq
 {\cal T}_a = \frac{1}{2}\left( p_x^2 + p_y^2 + V_a\,p_z^2 \right) \, , \qquad    V_a =  \frac 12\,k_1 (x^2 + y^2) + \frac{k_2}{x^2}  + \frac{k_3}{y^2}\,  ,
\label{ea}
\eeq
 is   HJ separable in Cartesian $(x,y,z)$  and  cylindrical  $(r,\phi,z)$ coordinates.  
This determines a superintegrable system endowed with   four independent constants of motion  given by $K_{a1}, K_{a2}, K_{a3}$  (\ref{da}) and $J_{a2}$ (\ref{db}). 
Furthermore,   $K_{a1}, K_{a2}, K_{a3}$ are mutually in involution and  $ {\cal T}_a  = \frac{1}{2}\bigl(K_{a2} +  K_{a3} \bigr)$.
\end{proposition} 
 
\subsect{Geodesic Hamiltonian ${\cal T}_b$ from   anisotropic oscillator}

Let $ {\cal T}_b$ be the 3D free Hamiltonian (\ref{CaHa})  with     $V_b$ (\ref{cc}). Since  $V_b$ is separable   in Cartesian and parabolic $(a,b)$ coordinates, we study the    separability of   $ {\cal T}_b $  in  Cartesian  $(x,y,z)$ and parabolic-cylindrical $(a,b,z)$ coordinates.

\subsubsect{ Cartesian separability}

  The HJ  equation yields
$$
\left(\fracpd{W}{x}\right)^2 + \left(\fracpd{W}{y}\right)^2  +  
V_b(x,y) \left(\fracpd{W}{z}\right)^2  = 2 E \, ,
$$
so that if we assume that $W$ can be written as $W = W_x(x) + W_y(y) + W_z(z)$ then we can perform a separation of variables obtaining the one-variable expressions 
$$
 (W_z')^2   = -\,\alpha  \,,{\quad\ } 
  (W_x')^2   - \alpha\left(2k_1 x^2 + k_3 x\right)  =  \beta +  E   \,,{\quad\ } 
 (W_y')^2   - \alpha\left( \frac 12\,k_1 y^2 + \frac{k_2}{y^2}\right) 
 =   -\beta +  E   \,,
$$
where $\alpha$ and $\beta$ are   two constants.
Each one of these expressions determines a constant of motion, namely,
\beq
 K_{b1}  = p_z  \,,{\qquad} 
 K_{b2}  = p_x^2 + \left(2 k_1x^2+k_3 x\right) p_z^2    \,,{\qquad} 
 K_{b3}  = p_y^2 + \left( \frac 12\,k_1y^2+\frac{k_2}{y^2}\right)p_z^2     \,,  
\label{dc}
\eeq
which, moreover, are  functionally independent   
$$
   \dd K_{b1}\,\wedge\, \dd K_{b2}\,\wedge\, \dd K_{b3}\ne 0 \,,{\quad} 
$$
and they satisfy the following properties
 $$
  \{K_{b1}, K_{b2}\} = 0 \,,{\qquad} 
  \{K_{b1},K_{b3}\} = 0 \,,{\qquad} \{K_{b2},K_{b3}\} = 0 \,,
  \qquad   {\cal T}_b  = \frac{1}{2}\bigl(K_{b2} +  K_{b3} \bigr) \,. 
$$

\subsubsect{Parabolic-cylindrical separability}

If we introduce the parabolic coordinates defined by
\beq
 x = \frac 12 \left(\pa^2 - \pb^2\right) \,,{\qquad} y =  \pa \pb \,, 
 \label{parab}
\eeq
the Hamiltonian ${\cal T}_b$  (\ref{CaHa}) and the potential $V_b$ (\ref{cc})  become 
$$
 {\cal T}_b = \frac{1}{2}\left(\frac{p_\pa^2+p_\pb^2}{ \pa^2+\pb^2} + V_b\,p_z^2\right) \,,{\quad\ }    
 V_b = \frac{1}{\pa^2+\pb^2}\left[\frac {k_1}2(\pa^6+\pb^6)  +  k_2 \left(\frac{1}{\pa^2}+\frac{1}{\pb^2}\right)  + \frac {k_3}2 (\pa^4-\pb^4)  \right]  \,, 
$$
and the HJ equation  adopts the form 
$$
  \frac{1}{\pa^2+\pb^2}\left[ \left(\fracpd{W}{\pa}\right)^2 + \left(\fracpd{W}{\pb}\right)^2 \right]+ V_b(\pa,\pb) \left(\fracpd{W}{z}\right)^2   =  2 E \,, 
$$
so that if we assume that $W$ is of the form $W = W_\pa(\pa) + W_\pb(\pb) + W_z(z)$ we can perform a separation of variables   obtaining first  $(W_z')^2   = -\,\gamma$ and then 
$$
  \left((W_\pa')^2  - 2 \pa^2 E  \right)+ \left((W_\pb')^2    - 2  \pb^2 E  \right)  = 
  \gamma \left( \frac {k_1}2\, \pa^6+ \frac{k_2}{\pa^2} + \frac {k_3}2\, \pa^4 \right) +   \gamma \left(\frac {k_1}2\, \pb^6+\frac{k_2}{\pb^2} - \frac {k_3}2\, \pb^4\right) \, ,
$$
providing three  integrals 
\begin{eqnarray}
&&J _{b1}=p_z\equiv  K_{b1} \, , \qquad
  J_{b2}  = p_\pa^2 + \left(\frac {k_1}2\, \pa^6+ \frac{k_2}{\pa^2} + \frac {k_3}2\, \pa^4\right)p_z^2   - 2 \pa^2 {\cal T}_b  \,, \nonumber\\
&& J_{b3}  = p_\pb^2 + \left(\frac {k_1}2\, \pb^6+ \frac{k_2}{\pb^2} -\frac {k_3}2\, \pb^4\right)p_z^2   - 2 \pb^2 {\cal T}_b  \,,  
\label{dd}
\end{eqnarray}
such that 
$\{J_{b1}\,,J_{b2}\} = 0$  and  $J_{b2} + J_{b3} = 0 $.

The following proposition summarizes these results. 

\begin{proposition}  \label{proposition2}
The 3D geodesic Hamiltonian given by
\beq
 {\cal T}_b = \frac{1}{2}\left( p_x^2 + p_y^2 + V_b\,p_z^2 \right) \,  \qquad  V_b = \frac 12 \, k_1 (4 x^2 + y^2) + \frac{k_2}{y^2}  + k_3 x \, ,
\label{eb}
\eeq 
is HJ separable in Cartesian $(x,y,z)$ and parabolic-cylindrical $(\pa,\pb,z)$ coordinates.
It represents a superintegrable system endowed with   four independent constants of motion corresponding to $K_{b1}, K_{b2}, K_{b3}$  (\ref{dc}) and $J_{b2}$ (\ref{dd}). Moreover,   $K_{b1}, K_{b2}, K_{b3}$ are mutually in involution and ${\cal T}_b  = \frac{1}{2}\bigl(K_{b2} +  K_{b3} \bigr)$.
\end{proposition}

\subsect{Geodesic Hamiltonian ${\cal T}_c$ from Kepler--Coulomb I}

Now we consider $ {\cal T}_c$   (\ref{CaHa})  with     $V_c$ (\ref{cd}). Recall that  $V_c$ is separable 
in polar and parabolic  coordinates, so that we analyse  the    separability of   $ {\cal T}_c$  in  cylindrical    and parabolic-cylindrical  coordinates. 

\subsubsect{Cylindrical separability}

In the variables $(r,\phi,z)$, the geodesic Hamiltonian $ {\cal T}_c$ is expressed as
$$
 {\cal T}_c = \frac{1}{2}\left( p_r^2 + \frac{p_\phi^2}{r^2} + V_c\,p_z^2 \right) \, , \qquad    V_c =  \frac{k_1}{r}  +  \frac{k_2}{r^2\sin^2\phi}  +  \frac{k_3 \cos\phi}{r^2\sin^2\phi}\,  .
 $$
The HJ equation reads
$$
   \left(\fracpd{W}{r}\right)^2 + \frac{1}{r^2}\left(\fracpd{W}{\phi}\right)^2 + 
   V_c(r,\phi) \left(\fracpd{W}{z}\right)^2   =  2 E \, .
$$
Hence if we assume that $W$ is of the form $W = W_r(r) + W_\phi(\phi) + W_z(z)$ we can perform a separation of variables, finding   first   that $(W_z')^2   = -\,\gamma$ and then 
$$
   r^2(W_r')^2   - 2 r^2 E  -  k_1 \gamma\,r =    -\,  (W_\phi')^2 
   + \gamma  \left(  \frac{k_2}{\sin^2\phi}  +  \frac{k_3 \cos\phi}{\sin^2\phi } \right) = \delta \, ,
$$
so that the following functions 
\begin{eqnarray}
 && K_{c1}  =  p_z  \,,\qquad
 K_{c2}  = p_\phi^2  +  \Bigl(  \frac{k_2}{\sin^2\phi}  +  \frac{k_3 \cos\phi}{\sin^2\phi } \Bigr)\,p_z^2\,,\nonumber\\[2pt] 
&& K_{c3}  = r^2 p_r^2 +  k_1r \,p_z^2   -  2\,r^2\,{\cal T}_c\, , \label{de}
\end{eqnarray}
are constants of motion such that 
$ \{K_{c1}\,,K_{c2}\} = 0$ and $K_{c2} +  K_{c3} = 0$.

\subsubsect{Parabolic-cylindrical separability}

 By introducing the parabolic coordinates  (\ref{parab}) we find that the Hamiltonian $ {\cal T}_c$  (\ref{CaHa})   is given by   
$$
{\cal T}_c = \frac{1}{2}\left(\frac{p_\pa^2+p_\pb^2}{ \pa^2+\pb^2} + V_c\,p_z^2\right) \,,{\qquad}    
V_c = \frac{1}{\pa^2+\pb^2}\left[2 k_1  +  k_2 \left(\frac{1}{\pa^2}+\frac{1}{\pb^2}\right)  
 +k_3 \left(\frac{1}{\pb^2}-\frac{1}{\pa^2}\right)   \right]  \,, 
\label{potc}
$$
so that the HJ equation becomes 
$$
  \frac{1}{\pa^2+\pb^2}\left[ \left(\fracpd{W}{\pa}\right)^2 + \left(\fracpd{W}{\pb}\right)^2 \right] + 
  V_c(\pa,\pb) \left(\fracpd{W}{z}\right)^2   =  2 E \, .
$$
By writing $W = W_\pa(\pa) + W_\pb(\pb) + W_z(z)$,   separability leads to 
$$
\left( (W_\pa')^2 - 2 \pa^2 E  \right) + \left( (W_\pb')^2   - 2  \pb^2 E  \right)  = 
  \gamma\, \left(   k_1 + \frac{k_2-k_3}{\pa^2} \right) 
  + \gamma\, \left(  k_1 + \frac{k_2+k_3}{\pb^2} \right)  \, .
$$
Therefore the following  functions 
\begin{eqnarray}
&& J_{c1}  =   p_z \equiv  K_{c1}   \,,\qquad
 J_{c2}  = p_\pa^2 + \left( k_1 + \frac{k_2-k_3}{\pa^2} \right) p_z^2 - 2 \pa^2\,{\cal T}_c  \,,\nonumber\\[2pt]
&& J_{c3} = p_\pb^2 + \left( k_1 + \frac{k_2+k_3}{\pb^2} \right) p_z^2  - 2 \pb^2\,{\cal T}_c \,,  
\label{df}
\end{eqnarray}
are three constants of motion fulfilling $ \{J_{c1}\,,J_{c2}\} = 0$ and $J_{c2} +  J_{c3} = 0$.

We conclude with the following statement.

\begin{proposition}  \label{proposition3}
The 3D geodesic Hamiltonian  
\beq
{\cal T}_c = \frac{1}{2}\left( p_x^2 + p_y^2 + V_c\,p_z^2\right)
\,,{\qquad} 
  V_c =  \frac{k_1}{\sqrt{x^2 + y^2}}  +  \frac{k_2}{y^2}  +  \frac{k_3x}{y^2 \sqrt{x^2 + y^2}} \,, 
\label{ec}
\eeq
is HJ separable in cylindrical $(r,\phi,z)$ and parabolic-cylindrical $(\pa,\pb,z)$ coordinates. 
This   is endowed with four independent constants of motion: the Hamiltonian itself, ${\cal T}_c$,  along with $K_{c1}, K_{c2}$ (\ref{de}) and $J_{c2}$ (\ref{df}).
The three integrals ${\cal T}_c, K_{c1}, K_{c2}$ are mutually in involution. 
\end{proposition}

\subsect{Geodesic Hamiltonian ${\cal T}_d$ from Kepler--Coulomb II}

Finally,  we consider  the four family  $ {\cal T}_d$   (\ref{CaHa})  with     $V_d$ (\ref{ce}). 
Since $V_d$ is separable in two types of parabolic coordinates, $(\pa,\pb)$ and $(\alpha,\beta)$, we study  the  separability of   $ {\cal T}_d$  in the corresponding  two types of  parabolic-cylindrical  coordinates.

\subsubsect{Parabolic-cylindrical separability I}

We   introduce   the parabolic coordinates $(\pa,\pb)$ (\ref{parab}) in   the  Hamiltonian $ {\cal T}_d$ yielding
$$
 {\cal T}_d = \frac{1}{2}\left(\frac{p_\pa^2+p_\pb^2}{ \pa^2+\pb^2} + V_d\,p_z^2\right) \,,{\qquad}    
 V_d =2\,\frac{k_1  +  k_2 \pa + k_3 \pb}{\pa^2+\pb^2}  \, .
$$
Hence the corresponding   HJ equation 
$$
  \frac{1}{\pa^2+\pb^2}\left[ \left(\fracpd{W}{\pa}\right)^2 + \left(\fracpd{W}{\pb}\right)^2 \right] + V_d(\pa,\pb) \left(\fracpd{W}{z}\right)^2   =  2 E
$$
admits separation of variables and it leads to  $(W_z')^2   = -\,\gamma$ and 
$$
 \left[ (W_\pa')^2 - 2 \pa^2 E  - \left(   k_1 + 2  k_2 \pa \right) \gamma \right]  + 
 \left[ (W_\pb')^2 - 2 \pb^2 E  - \left(   k_1 + 2  k_3 \pb \right) \gamma\right]  =  0   \,. 
$$
Therefore,  the following functions
\begin{eqnarray}
&&  K_{d1}  =   p_z \,,  \qquad
 K_{d2}  = p_\pa^2 + \left(   k_1 +  2 k_2 \pa \right) p_z^2 - 2 \pa^2 {\cal T}_d  \,,\nonumber\\[2pt] 
&& K_{d3}  = p_\pb^2 + \left(   k_1 + 2 k_3 \pb \right) p_z^2 - 2 \pb^2 {\cal T}_d \,,  
\label{dg}
\end{eqnarray}
are constants of motion  satisfying 
$\{K_{d1}\,,K_{d2}\} = 0$  and  $K_{d2}+  K_{d3} = 0$. 

\subsubsect{Parabolic-cylindrical separability II}

We  consider a second system of parabolic coordinates $(\al,\be) $ by rotating the original one $(\pa,\pb)$  (\ref{parab})  in the form
\beq
 \pa = \frac{1}{\sqrt{2}}(\al+\be) \, , \qquad  \pb = \frac{1}{\sqrt{2}}(\al-\be) \,,
  \label{parabb}
\eeq
in such a way that the Hamiltonian ${\cal T}_d$ is now given by
$$
 {\cal T}_d = \frac{1}{2}\left(\frac{p_\al^2+p_\be^2}{ \al^2+\be^2} + V_d\,p_z^2\right) \,,{\qquad}    
 V_d =\frac{2 k_1  +  k_2 \sqrt{2}(\al+\be)  + k_3 \sqrt{2}(\al-\be)}{\al^2+\be^2} \, .
$$
Thus  ${\cal T}_d$  determines the HJ equation
$$
  \frac{1}{\al^2+\be^2}\left[ \left(\fracpd{W}{\al}\right)^2 + \left(\fracpd{W}{\be}\right)^2 \right] +V_d(\al,\be) \left(\fracpd{W}{z}\right)^2   =  2 E
$$
 that also admits separability leading to three integrals
 \begin{eqnarray}
&&  J_{d1}  =   p_z \equiv   K_{d1} \,,  \qquad
 J_{d2}  = p_\al^2 + \left(   k_1 + \sqrt{ 2} ( k_2+k_3)  \al \right) p_z^2 - 2 \al^2 {\cal T}_d  \,,\nonumber\\[2pt] 
&& J_{d3}  = p_\be^2 + \left(   k_1 + \sqrt{ 2} ( k_2-k_3)  \be \right) p_z^2 - 2 \be^2 {\cal T}_d  \,,  
\label{dh}
\end{eqnarray}
such that 
$\{J_{d1}\,,J_{d2}\} = 0$  and  $J_{d2}+  J_{d3} = 0$.

These results are summarized as follows.

\begin{proposition}  \label{proposition4}
The 3D geodesic Hamiltonian given by 
\begin{equation}    
  {\cal T}_d = \frac{1}{2}\,\Bigl( p_x^2 + p_y^2 + V_d\,p_z^2\Bigr)      \,,{\quad\ } 
V_d =   \frac{k_1}{r}  +  k_2 \frac{\,\sqrt{r + x}\,}{r}  +  k_3 \frac{\,\sqrt{r - x}\,}{r}   
  \,,{\quad}  r^2 = x^2+y^2\,,    
    \label{ed}
\end{equation}     
is HJ separable in two sets of parabolic-cylindrical  coordinates $(\pa,\pb,z)$ and   $(\al,\be,z)$ which are related by a rotation. 
This system is  endowed with four  independent  constants of motion:  the Hamiltonian  ${\cal T}_d$, together with the   $K_{d1}, K_{d2}$ (\ref{dg}) and $J_{d2}$ (\ref{dh}).
The three integrals ${\cal T}_d, K_{d1}, K_{d2}$ are mutually in involution. 
\end{proposition} 

Notice that $K_{d2}$ and $J_{d2}$ can be written in the first set of parabolic-cylindrical  coordinates  $(\pa,\pb,z)$ as 
\begin{eqnarray}
&&\!\!\!\! \!\!\!\!  K_{d2}  =   \frac{\pb^2 p_\pa^2-\pa^2 p_\pb^2 }{\pa^2 + \pb^2} + \left(
 \frac{k_1( \pb^2-\pa^2 ) + 2 k_2 \pa \pb^2  - 2 k_3   \pa^2 b }{\pa^2 + \pb^2} \right) p_z^2  \,, \nonumber\\[2pt] 
&& \!\!\!\! \!\!\!\!   J_{d2}  =  \frac{(\pa  p_\pb -\pb  p_\pa )(\pa  p_\pa-\pb  p_\pb)}{\pa^2 + \pb^2} - 
\left( \frac{2 k_1 \pa \pb  + k_2 (\pa^2 - \pb^2) \pb  - k_3  (\pa^2 - \pb^2)  \pa}{\pa^2 + \pb^2} \right) p_z^2 \, ,
\label{xa}
\end{eqnarray}
and they can be   interpreted as generalised versions of the  two Laplace--Runge--Lenz constants of motion of the 2D KC  problem.

\sect{Hamiltonians ${\cal H}$ endowed with multiple separability on 3D curved spaces }

The next step in our approach is to add a 3D potential $\pot_r$ to each  superintegrable geodesic Hamiltonian $ {\cal T}_r $ constructed in the previous section,  thus leading to a natural   Hamiltonian $ {\cal H}_r$ in the form
\beq
{\cal H}_r= {\cal T}_r + \pot_r= \frac{1}{2}\left( p_x^2 + p_y^2 + V_r(x,y)\,p_z^2 \right)+\pot_r(x,y,z) \, , \qquad r=a,b,c,d\, . 
  \label{CaHamX}
\eeq
We  now  require  the complete Hamiltonian ${\cal H}_r$ to be HJ separable in the same two sets of coordinates as its kinetic component. Therefore, due to the structure of   ${\cal T}_r $, we   assume that the 3D potential is  given by
\beq
\pot_r(x,y,z)=U_r(x,y) +  V_r (x,y) Z(z) \, , \qquad r=a,b,c,d\, ,
\label{CaHamXb}
\eeq
where $U_r(x,y)$ is  a function  to be determined for each family,  $V_r (x,y)$ is just the known 2D potential,  and $ Z(z)$ is  always an arbitrary smooth function for the four families.
 Hence the generic initial Hamiltonian reads
\begin{equation}   
{\cal H}_r= {\cal T}_r + \pot_r= \frac{1}{2}\left( p_x^2 + p_y^2 + V_r(x,y)\,p_z^2 \right)+U_r(x,y) +  V_r (x,y) Z(z)\, , \qquad r=a,b,c,d\, . 
  \label{CaHam}
\end{equation} 
Then  multiseparability will give rise to the compatible explicit form  for $U_r(x,y)$   together with four independent constants of motion. We remark that, for the four families,  $U_r$ will be formally similar to $ V_r$ but with different coefficients $t_i$ instead of $k_i$ $(i=1,2,3)$.
Consequently, the resulting  Hamiltonian ${\cal H}_r$ will always determine a superintegrable system but, in general, not a maximally superintegrable one.

Since computations are quite similar to the previous ones, we shall omit most technical details and only provide the main results.

\subsect{Hamiltonian ${\CaH}_a$  from   isotropic oscillator}

Let us consider ${\cal H}_a$ (\ref{CaHam}) with  ${\cal T}_a$  given in (\ref{ea}) and impose that such a system  
preserve the same multiple separability studied in Section 4.1. 
 In this case,   separability in  Cartesian and  cylindrical  coordinates implies that  
  $$
  U_a(x,y) = A(x) + B(y) \, , \qquad U_a(r,\phi)= F(r) +  \frac{G(\phi)}{r^2} \, .
  $$
  These two restrictions determine the form of the potential $U_a$ through the above functions   that turn out to be
 \bea
&&
   A(x) =   \frac{1}{2}\,t_1 x^2  +  \frac{t_2}{x^2}  \,,{\qquad}
 B(y)  =   \frac{1}{2}\,t_1 y^2  +  \frac{t_3}{y^2}  \,, \nonumber\\[2pt]
 &&F(r) = \frac 12\, t_1 r^2 \, , \qquad G(\phi)= \frac{ t_2}{\cos^2\phi}  +  \frac{ t_3}{\sin^2\phi}  \, ,
 \nonumber 
 \eea
  where, from now on, $t_1,t_2,t_3$  denote three  arbitrary real constants. Thus  $U_a$ is a function formally similar to $V_a$ but with different coefficients.  

The final results are summarized in the following statement which generalises Proposition~\ref{proposition1}.

\begin{proposition}  \label{proposition5}
The   3D Hamiltonian ${\cal H}_a$  (\ref{CaHam}) with $V_a$ (\ref{ea}) and similar $U_a$ with coefficients $t_i$
is HJ separable in Cartesian $(x,y,z)$ and cylindrical  $(r,\phi,z)$ coordinates  and  it is endowed with    four independent quadratic constants of motion given by
\begin{eqnarray*}
&& {\cal K}_{a1}= p_z^2 + 2 Z(z)  \,,  \nonumber\\[2pt]
&& {\cal K}_{a2} = p_x^2 +\left( \frac 12\, k_1x^2+\frac{k_2}{x^2}\right)\left(p_z^2  + 2 Z(z)\right)+   t_1 x^2  +  \frac{2t_2}{x^2}   , \nonumber\\[2pt]
&& {\cal K}_{a3} = p_y^2 +\left( \frac 12\,  k_1y^2+\frac{k_3}{y^2}\right)\left(p_z^2  + 2 Z(z)\right)   + t_1 y^2  +  \frac{2t_3}{y^2}   ,  \nonumber\\[2pt]
&& {\cal J}_{a2}= p_\phi^2  +\left(   \frac{k_2}{\cos^2\phi}  +  \frac{k_3}{\sin^2\phi}\right)  \left(p_z^2  + 2 Z(z)\right) +    \frac{2t_2}{\cos^2\phi}  +  \frac{2t_3}{\sin^2\phi}  . 
\end{eqnarray*}
The three integrals ${\cal K}_{a1}, {\cal K}_{a2}, {\cal K}_{a3}$ are mutually in involution and  $ {\cal H}_a  = \frac{1}{2}\bigl({\cal K}_{a2} +  {\cal K}_{a3} \bigr)$.\end{proposition}

\subsect{Hamiltonian ${\CaH}_b$  from   anisotropic oscillator}

Let ${\cal H}_b$  be the Hamiltonian   (\ref{CaHam}) with  kinetic term ${\cal T}_b$  given in (\ref{eb}). 
According to Section 4.2, we   impose separability in both Cartesian and parabolic-cylindrical  coordinates  which means that 
$$
U_b(x,y) = A(x) + B(y) \, , \qquad
U_b(\pa,\pb) = \frac{C(\pa) + D(\pb)}{\pa^2+\pb^2} \, ,
$$
 leading to
\bea
&& A(x) =   2 t_1 x^2  +  t_3 x \,,{\qquad}
 B(y)  =   \frac{1}{2}\,t_1 y^2  +  \frac{t_2}{y^2}  \,, \nonumber\\[2pt]
 &&
 C(\pa)  = \frac{ t_1}{2}\, \pa^6+ \frac{t_2}{\pa^2} +\frac{ t_3}{2}\, \pa^4  \, ,\qquad  D(\pb)  = \frac{ t_1}{2}\, \pb^6+ \frac{t_2}{\pb^2} -\frac{ t_3}{2}\, \pb^4   \, .
\nonumber
\eea
Hence  $U_b$ is again a function formally similar to $V_b$.

The final  results  generalise those achieved in  Proposition~\ref{proposition2} as follows.

\begin{proposition}  \label{proposition6}
The   3D Hamiltonian ${\cal H}_b$  (\ref{CaHam}) with $V_b$ (\ref{eb}) and similar $U_b$ with coefficients $t_i$
is   HJ separable in Cartesian $(x,y,z)$ and parabolic-cylindrical $(\pa,\pb,z)$ coordinates. This is endowed with the following  four independent quadratic constants of motion 
\begin{eqnarray*}
&& {\cal K}_{b1}= p_z^2  +  2 Z(z)\,,\\[2pt]
&&  {\cal K}_{b2}  = p_x^2  + \left(2 k_1x^2+k_3 x\right) \bigl(p_z^2  +  2 Z(z)\bigr)  +  4 t_1 x^2  + 2 t_3 x   \,, \\[2pt]
&&  {\cal K}_{b3} =  p_y^2  + \left( \frac 12\,k_1y^2 + \frac{k_2}{y^2}\right)\bigl(p_z^2  +  2 Z(z)\bigr) +   t_1 y^2  +  \frac{2 t_2}{y^2} \,, \\[2pt]
&&  J_{b2}  = p_\pa^2 + \left( \frac{k_1}2\, \pa^6+ \frac{k_2}{\pa^2} + \frac{k_3}2\, \pa^4\right)\bigl(p_z^2  +  2 Z(z)\bigr) +  { t_1} \pa^6+ \frac{2t_2}{\pa^2} + { t_3} \pa^4  - 2 \pa^2  {\CaH}_b \,, 
\end{eqnarray*}
such that $ {\cal K}_{b1},  {\cal K}_{b2},  {\cal K}_{b3}$ are mutually in involution and ${\cal H}_b  = \frac{1}{2}\bigl( {\cal K}_{b2} +   {\cal K}_{b3} \bigr)$.
\end{proposition}

\subsect{Hamiltonian ${\CaH}_c$  from Kepler--Coulomb I}

 Now we consider ${\cal H}_c$ (\ref{CaHam}) with  ${\cal T}_c$   (\ref{ec}) and impose  the separability  in cylindrical and parabolic-cylindrical coordinates as   in Section 4.3. 
This means that $U_c$ must admit the following expressions 
$$
U_c(r,\phi) =  F(r) +  \frac{G(\phi)}{r^2} \,,{\qquad}  U_c(\pa,\pb) = \frac{C(\pa) + D(\pb)}{\pa^2+\pb^2} \, .
$$
Compatibility among these two separabilities leads to
\bea
&& F(r)= \frac{t_1}{r}  \, , \qquad   G(\phi)  =  \frac{t_2}{\sin^2\phi}  +  \frac{t_3\cos\phi}{\sin^2\phi}  \, ,\nonumber\\[2pt]
&& C(\pa)=    t_1 + \frac{ t_2-t_3}{\pa^2}   \,,{\qquad}
  D(\pb)  =    t_1 + \frac{t_2+ t_3}{\pb^2}   \, .
 \nonumber
\eea
Notice that $U_c$  is formally similar to $V_c$.
Then we   conclude with the following statement (to be compared with Proposition~\ref{proposition3}).

\begin{proposition}  \label{proposition7} 
The   3D Hamiltonian ${\cal H}_c$  (\ref{CaHam}) with $V_c$ (\ref{ec}) and similar $U_c$ with coefficients $t_i$
is  HJ separable in cylindrical $(r,\phi,z)$ and parabolic-cylindrical $(\pa,\pb,z)$ coordinates. 
This  system is endowed with four  independent integrals,  which are  the Hamiltonian itself,  ${\CaH}_c$, along with   
\begin{eqnarray*}
&& {\cal K}_{c1} = p_z^2 + 2 Z(z)  \,,  \\[2pt]
&&   {\cal K}_{c2}  = p_\phi^2 +\left(   \frac{k_2}{\sin^2\phi}  +  \frac{k_3\cos\phi}{\sin^2\phi}\right)  \bigl(p_z^2  + 2 Z(z)\bigr)  +  \frac{2t_2}{\sin^2\phi}  +  \frac{2t_3\cos\phi}{\sin^2\phi}   \,,\\[2pt]
&&   {\cal J}_{c2} = p_\pa^2 + \Bigr(  k_1 + \frac{k_2-k_3}{\pa^2} \Bigl) \bigl(p_z^2  + 2 Z(z)\bigr)  +   2t_1 + \frac{2(t_2-t_3)}{\pa^2}  - 2 \pa^2 {\CaH}_c  \,,
  \end{eqnarray*}
such that the three constants of motion ${\cal H}_c, {\cal K}_{c1}, {\cal K}_{c2}$ are mutually in involution. 
\end{proposition} 

\subsection{Hamiltonian ${\CaH}_d$ from Kepler--Coulomb II}

 Finally,   we consider the fourth family ${\cal H}_d$ (\ref{CaHam}) with  ${\cal T}_d$   (\ref{ed}) and require to preserve  the 
  separability  in  parabolic-cylindrical coordinates of type I (\ref{parab}) and type II (\ref{parabb})  similarly to     Section 4.4. 
Hence multiple separability means that $U_d$ must admit the following expressions 
$$
U_d(\pa,\pb) =  \frac{C(\pa) + D(\pb)}{\pa^2+\pb^2} \,,{\qquad}  U_d(\al,\be) = \frac{L(\al) + M(\be)}{\al^2+\be^2} \, ,
$$
which yields
\bea
&& C(\pa)=t_1+2t_2 \pa\, ,\qquad D(\pb)=t_1+2t_3 \pb\, ,\nonumber\\[2pt]
&&L(\al)= t_1+\sqrt{2}(t_2+t_3)\al \, ,\qquad M(\be)= t_1+\sqrt{2}(t_2-t_3)\be  \, ,
\nonumber
\eea
providing a potential $U_d$ formally similar to $V_d$. 
The final results, that generalise those given in Proposition~\ref{proposition4},  are summarized as follows.

\begin{proposition}\label{proposition8}  
The 3D Hamiltonian ${\cal H}_d$  (\ref{CaHam}) with $V_d$ (\ref{ed}) and similar $U_d$ with coefficients $t_i$ is HJ separable in two types of parabolic-cylindrical coordinates: $(\pa,\pb,z)$ and   $(\al,\be,z)$.   This is  endowed with four independent constants of motion: the Hamiltonian itself ${\CaH}_d$   together with the following  three  functions 
\begin{eqnarray*}
&& {\cal K}_{d1} = p_z^2 + 2 Z(z)  \,,  \\[2pt]
&&   {\cal K}_{d2}  = p_\pa^2 + \left(   k_1 +  2 k_2 \pa \right) \bigl(p_z^2  + 2 Z(z)\bigr) +  2t_1 +  4 t_2 \pa - 2 \pa^2 {\cal H}_d   \,,\\[2pt]
&&   {\cal J}_{d2} =  p_\al^2 + \left(   k_1 + \sqrt{ 2} ( k_2+k_3)  \al \right) \bigl(p_z^2  + 2 Z(z)\bigr) +2t_1+2\sqrt{2}(t_2+t_3)\al  - 2 \al^2 {\cal H}_d  \, .
  \end{eqnarray*}
The three integrals ${\cal H}_d, {\cal K}_{d1}, {\cal K}_{d2}$ are mutually in involution. 
\end{proposition}

In   parabolic-cylindrical coordinates $(\pa,\pb,z)$, the integrals ${\cal K}_{d2}$ and ${\cal J}_{d2}$ can be rewritten as 
\begin{eqnarray}
&&\!\!\!\!\!\!\!\!\!\!\!\!
 {\cal  K}_{d2}  =   \frac{\pb^2 p_\pa^2-\pa^2 p_\pb^2 }{\pa^2 + \pb^2} + \left(
 \frac{k_1( \pb^2-\pa^2 ) + 2 k_2 \pa \pb^2  - 2 k_3   \pa^2 \pb }{\pa^2 + \pb^2} \right) \bigl(p_z^2 + 2 Z(z)\bigr)     \nonumber\\[2pt] 
 &&\qquad\qquad + \frac{2t_1( \pb^2-\pa^2 ) + 4t_2 \pa \pb^2  - 4 t_3   \pa^2 \pb }{\pa^2 + \pb^2}
  \,, \nonumber\\[2pt] 
&&\!\!\!\!\!\!\!\!\!\!\!\!
 {\cal  J}_{d2}  =  \frac{(\pa  p_\pb -\pb  p_\pa )(\pa  p_\pa-\pb  p_\pb)}{\pa^2 + \pb^2} - 
\left( \frac{2 k_1 \pa \pb  + k_2 (\pa^2 - \pb^2) \pb  - k_3  (\pa^2 - \pb^2)  \pa}{\pa^2 + \pb^2} \right) \bigl(p_z^2 + 2 Z(z)\bigr)   \nonumber\\[2pt] 
&& \qquad\qquad - \frac{4 t_1 \pa \pb  + 2t_2 (\pa^2 - \pb^2) \pb  - 2t_3  (\pa^2 - \pb^2)  \pa}{\pa^2 + \pb^2}
 \, ,
\label{xb}
\end{eqnarray}
which, as we shall see later on, can be  regarded  as the generalised counterpart  of the  2D Laplace--Runge--Lenz  vector corresponding to the 2D KC  system.

\subsection{Comments}\label{section55}

So far,  by applying the Eisenhart formalism, we have achieved the extension sequence:
\beq
H_r=T+V_r\ \longrightarrow \  {\cal T}_r\ \longrightarrow \    {\cal H}_r= {\cal T}_r+\pot_r \, , \qquad r=a,b,c,d \, .
\label{seq}
\eeq
Obviously,  the reverse process, firstly,  corresponds to  set $\pot_r \equiv 0$, that is, $t_i=0$ $(i=1,2,3)$ and $Z(z)\equiv0$,  so that ${\cal H}_r\to  {\cal T}_r$ and the integrals ${\cal K}_{ri}\to { K}_{ri}$,   ${\cal J}_{ri}\to { J}_{ri}$;  thus Propositions 5--8 reduce to Propositions 1--4, respectively. 
  And secondly, to set $p_z$ constant, say $p_z=\sqrt{2}$,   gives ${\cal T}_r\to  {H}_r = T+V_r$ in the form described in Section 3.

Now we comment on some   characteristics of these new four families of 3D superintegrable Hamiltonians endowed with four independent constants of motion.

The four geodesic Hamiltonians ${\cal T}_r$  are endowed with an exact Noether symmetry since they are invariant under translation along the $z$-axis and the Noether theorem states the conservation of the momentum $p_z$, which is just their common integral $K_{r1}$. 
The remaining three constants of motion are quadratic and homogeneous in the momenta. 
The coefficients of such integrals can be considered as the components of Killing tensors of the underlying 3D Riemannian metric (\ref{ba}).  In this respect,  let us  recall the
 relation of Killing tensors with   Hamiltonian dynamics.

A Killing tensor ${\bf K}$ of valence   $p$ defined in a Riemannian manifold $(M, g)$  is a symmetric $(p, 0)$ tensor satisfying the Killing tensor equation \cite{Thom86}--\cite{RajaratMcLen14}
\begin{equation}
  [{\bf K},g]_S = 0 \,,  \label{KillingEquation}
\end{equation}
where $[\cdot,\cdot]_S$ denotes the Schouten bracket (bilinear operator representing the natural generalisation of the Lie bracket of vector fields).  When $p=1$ the Killing tensor 
reduces to a Killing vector $X\in{ \mathfrak{X}}(M)$ (generator of isometries),  the bracket becomes a Lie derivative and the Killing equation reduces to $\CaL_X(g)=0$.   
When $p=2$ the metric tensor $g$ is itself a trivial Killing tensor.
The set ${\CaK}^p(M)$ of all the Killing  tensors of valence $p$ on $M$ is a vector space. 
If $M$ is a space of constant curvature  then the dimension  $d$ of ${\CaK}^p(M)$ is giving by the Delong--Takeuchi--Thompson formula \cite{ChadGMc06}
 (that generalises the expression $\frac 12n(n+1)$ for Killing vectors)
$$
  d = \dim {\CaK}^p(M) = \frac{1}{n} {n+p \choose p+1}{n+p-1 \choose p}
  \,,\quad p\ge 1  \,. 
$$
In the more general case of a  space of nonconstant curvature,  the dimension of the vector space is a value lower than
 $d$. 

We are now interested in the particular case with $p=2$. In this case, the Killing tensor ${\bf K}$ determines  a  homogeneous quadratic function $F_K = K^{ij}p_i p_j$ and then the Killing equation can be rewritten as the vanishing of the Poisson bracket of two functions 
$$
 \{ K^{ij}p_i p_j  \,,\,  g^{ij}p_i p_j\} = 0 \,. 
$$
This means that the function $F_K$, associated to the tensor $\bf K$,  is a first integral of the geodesic flow determined by the Hamiltonian ${\cal T} = \frac 12 g^{ij}p_i p_j$.

The four families of 3D geodesic Hamiltonian systems, ${\cal T}_r$  ($r =a,b,c,d$),  are determined by metric tensors $g_r$  given by 
$$
  g_r^{ij} = {\rm diag}(1,1,V_r) \,,{\quad} r =a,b,c,d\, , 
$$
and the result is that the  four configuration spaces $(\mathbb{R}^3,g_r)$ are endowed with a Killing vector $X=\partial/\partial z$ (determining the linear constant $p_z$) and three $p=2$ Killing tensors determining the three quadratic integrals of motion. 
 
   We must mention that there  exist systems with higher-order constants of motion that, in differential geometric terms,  are related with the existence of $p>2$ Killing tensors.  
For example, a cubic integral of motion \cite{Grav04Jmp}--\cite{CCR13JPa} means that the configuration space admits a nontrivial symmetric $(3, 0)$ tensor satisfying the Killing tensor equation (\ref{KillingEquation}) and determining a function $F_K = K^{ijk}p_i p_j p_k$ satisfying 
$$
 \{ K^{ijk}p_i p_j p_k \,,\,  g^{ij}p_i p_j\} = 0 \,,
$$  
and representing a first-integral for the geodesic motion (the existence of higher-order Killing tensors for systems  in external gauge fields is analyzed in \cite{vHolt07, Visien10}). 
Ne\-vertheless we restrict our study to the two above mentioned cases:  Killing vector fields and $p=2$  Killing tensors.   

Another interesting property deserving to be  mentioned is the close relation of the two integrals  $({\cal K}_{d2}, {\cal J}_{d2})$ (\ref{xb})  of ${\cal H}_d$ (and also $(K_{d2},J_{d2})$   (\ref{xa}) for ${\cal T}_d$)  with the Laplace--Runge--Lenz 2-vector since their first term   in parabolic  $(\pa,\pb)$ and Cartesian $(x,y)$ coordinates reads as
\bea
&& {\cal K}_{d2} \  :\    \frac{\pb^2 p_\pa^2-\pa^2 p_\pb^2 }{\pa^2 + \pb^2}= -2 (x p_y-y p_x) p_y \, ,\nonumber\\[2pt]
&&
 {\cal  J}_{d2}  \  :\    \frac{(\pa  p_\pb -\pb  p_\pa )(\pa  p_\pa-\pb  p_\pb)}{\pa^2 + \pb^2} =2(x p_y-y p_x) p_x \, .
\nonumber
\eea
As it is well known the existence of this conserved vector is one of the main characteristics of the KC problem and the importance of this fact have led to the study of systems  admitting generalisations of the Laplace--Runge--Lenz vector \cite{ComRunge},  \cite{Redm64}--\cite{Nikitin14}. 
 In particular,  the   Hamiltonian ${\cal H}_d$,  defined on a 3D curved space of nonconstant curvature,  could be considered as a new cornerstone in order to  generalised  the KC problem and consequently $ {\cal K}_{d2}$ and $ {\cal J}_{d2}$ could be considered as new ways of representing the  Laplace--Runge--Lenz vector;  clearly, this property also holds for   the geodesic Hamiltonian ${\cal T}_d$.

\newpage

\sect{3D geodesic Hamiltonians $\wt{\cal T}$ with a position-dependent mass}
\label{section6}

The   sequence (\ref{seq}) can further be enlarged through the introduction of an `appropriate' position-dependent mass (PDM) \cite{Vak05}--\cite{Quesne15Jmp}, in such a manner that new (generalised) Hamiltonians can be obtained by requiring once again to preserve separability/superintegrabilty.

More specifically, let us consider the Euclidean plane  with metric $\dd s^2$, free Lagrangian $L$  and geodesic Hamiltonian $T$ in  Cartesian coordinates. The introduction of a  PDM,  $\mu(x,y)$, determines a metric $\dd s_\mu^2$ on a 2D Riemannian space (generally,  of nonconstant curvature)  with associated free Lagrangian $L_\mu$ and  geodesic Hamiltonian  $T_\mu$   given by
$$
 \dd s_\mu^2 =\mu\,   \dd s^2=\mu \bigl( \dd x^2 +  \dd y^2\bigr) \,,{\quad\ }\, 
  L_\mu =\mu\, L= \frac 12\,\mu \bigl( v_x^2 +  v_y^2\bigr)\,,{\quad\ }\, 
  T_\mu =\frac 1{\mu}\, T=  \frac 1{2\mu}\,\bigl( p_x^2 +  p_y^2\bigr)  \,,
$$
i.e. the new metric is conformally Euclidean. 

Therefore in the Eisenhart formalism we are considering, which starting from the Euclidean plane with a potential $V_r(x,y)$  leads to a geodesic motion in a  3D Riemannian configuration space, we  can express the new  metric $\dd \wt\sigma_{r}^2$ and   the new geodesic Hamiltonian $\wt{\cal T}_r$ as    (see (\ref{ba}) and (\ref{CaHa}))
\bea
&&  \dd \wt\sigma_r^2 =\mu_r\,  \dd \sigma_r^2= \mu_r(x,y) \left( \dd x^2 +  \dd y^2 + \frac{\dd z^2}{V_r(x,y)} \right) , 
\nonumber\\[2pt]
&&
 \wt{\cal T}_r = \frac 1{\mu_r} \,   {\cal T}_r = \frac{1}{2\mu_r(x,y)}\left( p_x^2 + p_y^2 + V_r(x,y)\,p_z^2 \right) , \qquad r=a,b,c,d\, . 
\label{hax}
\eea

The new geodesic dynamics  determined by $\wt{\cal T}_r $  must be a deformation of the initial one provided by ${\cal T}_r$ \cite{Ra15Jmp} in the sense that the PDM  $\mu_r$, and so $\wt{\cal T}_r $, will depend on a real parameter $\kp$ in such a way that the following properties must be satisfied:

\begin{itemize}
\item[(i)] The PDM $\mu_r(\kp)$ must preserve the multiple separability of the original dynamics as established in Section 4. 
\item[(ii)]   The new geodesic Hamiltonian $\wt{\cal T}_r(\kp)$  must be a continuous function of $\kp$ (in a certain domain of the parameter). 
\item[(iii)]  When taking the limit $\kp\to 0$ the PDM must satisfy $\mu_r(\kp)\to 1$, so that the dynamics of the original geodesic Hamiltonian  ${\cal T}_r$ is recovered.
\end{itemize}

In what follows we study the separability and superintegrability of the   geodesic Hamiltonians $\wt {\cal T}_r(\kp)$  obtained by introducing an adequate  $\kp$-PDM  $\mu_r(\kp)$ in the four families of geodesic Hamiltonians ${\cal T}_r$  ($r=a,b,c,d$) described in Section 4,   fulfilling the three above requirements. 
We remark that the explicit form for $\mu_r$ is, in fact, determined by the `principal' potential within each family, that is, the $k_1$-term.

\subsection{Geodesic Hamiltonian $\wt{{\cal T}}_a$  with PDM from   isotropic oscillator}

Let us consider the geodesic Hamiltonian  $\wt{{\cal T}}_a$ (\ref{hax}) with ${{\cal T}}_a$ given in (\ref{ea}) and PDM defined by
\beq
   \mu_a(x,y) = 1 - \kp\,r^2 \,,{\qquad} 
 r^2 = x^2+y^2 \,. 
\label{mass1}
\eeq
The parameter $\kp$ can take both positive and negative values. If $\kp<0$,  the  dynamics from  $\wt{{\cal T}}_a$  is correctly defined for all the values of the variables; 
nevertheless, when $\kp>0$, the Hamiltonian (and the associated  dynamics)  has a singularity at $1 -\,\kp\,r^2=0$, so in this case the  dynamics is only defined in the interior of the circle with radius $r=1/\sqrt{\kp}$,   that is, the region in which $\wt{{\cal T}}_a$ is positive definite.

The HJ   equation  in Cartesian coordinates  takes the form 
$$
 \frac{1}{1  - {\kp} (x^2+y^2)} \left[  \left(\fracpd{W}{x}\right)^2 + \left(\fracpd{W}{y}\right)^2  
 +  V_a(x,y) \left(\fracpd{W}{z}\right)^2  \right]   =2 E  \,,
$$
and admits separability giving rise to three independent first integrals 
\bea
&& \wt{K}_{a1}   = p_z  \,, \qquad 
\wt{K}_{a2}  = p_x^2 + \left({\frac 12}\, k_1x^2+\frac{k_2}{x^2}\right)p_z^2 +  2 {\kp}\,x^2\, \wt{{\cal T}}_a  \,,\nonumber\\[2pt] 
&&  \wt{K}_{a3}=   p_y^2 + \left( {\frac 12}\, k_1y^2+\frac{k_3}{y^2}\right)p_z^2  + 2 {\kp}\,y^2\, \wt{{\cal T}}_a  \, . 
\label{ga}
\eea
Meanwhile,  the  HJ  equation  in cylindrical  coordinates 
$$
 \frac{1}{1 - \kp\,r^2}  \left[ \left(\fracpd{W}{r}\right)^2 + \frac{1}{r^2}\left(\fracpd{W}{\phi}\right)^2 
  +  V_a(r,\phi) \left(\fracpd{W}{z}\right)^2 \right]=  2E
$$
also admits separability providing a fourth functionally independent constant of motion
\beq
\wt J_{a2} = p_\phi^2  +  \left(   \frac{k_2}{\cos^2\phi}  +  \frac{k_3}{\sin^2\phi}\right) p_z^2 
    \,. 
\label{gb}
\eeq
 Consequently, we find:
  
\begin{proposition}  \label{proposition9}
The 3D $\kp$-dependent geodesic Hamiltonian  $\wt{\cal T}_a(\kp)$ (\ref{hax}), with   ${{\cal T}}_a$  (\ref{ea}) and PDM (\ref{mass1}),  
$$
  \wt{\cal T}_a = \frac 1{\mu_a} \,   {\cal T}_a = \frac{1}{2\mu_a(x,y)}\left( p_x^2 + p_y^2 + V_a(x,y)\,p_z^2 \right) , {\qquad}  \mu_a(x,y) = 1 - \kp\,r^2  \, , 
$$ 
is HJ separable in Cartesian $(x,y,z)$ and  cylindrical  $(r,\phi,z)$ coordinates. 
This  is endowed with  four independent constants of motion  given by $\wt K_{a1}, \wt K_{a2}, \wt K_{a3}$  (\ref{ga}) and $\wt J_{a2}$ (\ref{gb}). 
Moreover,  $\wt K_{a1}, \wt K_{a2}, \wt K_{a3}$ are mutually in involution and  $ \wt{\cal T}_a  = \frac{1}{2}\bigl(\wt K_{a2} +  \wt K_{a3} \bigr)$.
\end{proposition} 

Notice that the integral  $\wt{J}_{a2}$   is $\kp$-independent and   coincides with the original one  $J_{a2}$ (\ref{db}). The two remaining constants of motion   satisfy  the   limits  
$ \wt{K}_{a2} \,\to\, K_{a2}$ and $ \wt{K}_{a3} \,\to\,  K_{a3}$ when ${\kp\to 0}$, so recovering  (\ref{da}) and 
Proposition \ref{proposition1}.

\subsection{Geodesic Hamiltonian $\wt{{\cal T}}_b$  with PDM from  anisotropic oscillator}

For the second family $\wt{{\cal T}}_b$ (\ref{hax}) with ${{\cal T}}_b$ given in (\ref{eb}) we define the PDM  by
\beq
  \mu_b(x,y) = 1 - \kp\,  x  \,. 
\label{mass2}
\eeq
Hence the region in which $  \mu_b(x,y)$  is positive definite   is given by  $x<1/\kp$. 
 
Next, the  HJ  equation   written in Cartesian coordinates reads
$$
   \frac{1}{1  - \kp x}  \left[ \left(\fracpd{W}{x}\right)^2 + \left(\fracpd{W}{y}\right)^2  +  
V_b(x,y) \left(\fracpd{W}{z}\right)^2  \right]   = 2 E  \,, 
$$
so it admits separability and leads to the following three independent  constants of motion
\begin{eqnarray}
&&  \wt K_{b1}  = p_z  \,,{\qquad} 
 \wt K_{b2}  = p_x^2 + \left(2 k_1x^2+k_3 x\right) p_z^2  + 2 {\kp} x  \wt{{\cal T}}_b   \,,  \nonumber\\[2pt]
&&\wt  K_{b3}  = p_y^2 + \left( \frac 12\,k_1y^2+\frac{k_2}{y^2}\right)p_z^2   \, .  
\label{gc}
\end{eqnarray}
 The  HJ equation when   written in parabolic coordinates $(\pa,\pb,z)$  (\ref{parab}),
$$
\frac{1}{1-\frac 12\kp(\pa^2-\pb^2)} \left\{  \frac{1}{\pa^2+\pb^2}\left[ \left(\fracpd{W}{\pa}\right)^2 + \left(\fracpd{W}{\pb}\right)^2 \right]+ V_b(\pa,\pb) \left(\fracpd{W}{z}\right)^2   \right\} =  2 E \, ,
$$
also admits separability, yielding to another functionally  independent constant of motion, namely 
\beq
 \wt{J}_{b2} =p_\pa^2 + \left(\frac {k_1}2\, \pa^6+ \frac{k_2}{\pa^2} + \frac {k_3}2\, \pa^4\right)p_z^2   -  \pa^2\left( 2-  \kp   \pa^2\right) 
\wt{{\cal T}}_b\,. 
\label{gd}
\eeq

Thus we conclude with the following statement.  

\begin{proposition}  \label{proposition10}
The 3D $\kp$-dependent geodesic Hamiltonian  $\wt{{\cal T}}_b(\kp)$ (\ref{hax}) with ${{\cal T}}_b$   (\ref{eb}) and PDM (\ref{mass2}) 
$$
  \wt{\cal T}_b = \frac 1{\mu_b} \,   {\cal T}_b = \frac{1}{2\mu_b(x,y)}\left( p_x^2 + p_y^2 + V_b(x,y)\,p_z^2 \right) , {\qquad}  \mu_b(x,y) = 1 - \kp\,x  \, , 
$$ 
 is HJ separable in Cartesian $(x,y,z)$ and parabolic-cylindrical $(\pa,\pb,z)$ coordinates. This    is endowed with    four independent constants of motion   corresponding to $\wt K_{b1}, \wt K_{b2}, \wt K_{b3}$  (\ref{gc}) and $\wt  J_{b2}$ (\ref{gd}). 
Furthermore,   $\wt  K_{b1}, \wt  K_{b2}, \wt  K_{b3}$ are mutually in involution and $\wt {\cal T}_b  = \frac{1}{2}\bigl(\wt K_{b2} +  \wt K_{b3} \bigr)$.
\end{proposition} 

Notice  that   the results given in Proposition 2 are straightforwardly recovered under taking the limit   $\kp\to 0$.

\subsection{Geodesic Hamiltonian $\wt{{\cal T}}_c$  with PDM from Kepler--Coulomb I}

Now  we consider the geodesic Hamiltonian  $\wt{{\cal T}}_c$ (\ref{hax}) with ${{\cal T}}_c$ given in (\ref{ec}) and with a  PDM defined by
\beq
  \mu_c(x,y) =    1 - \frac{\kp}{r} \,,{\qquad} 
 r^2 = x^2+y^2 \,. 
\label{mass3}
\eeq
Hence, if  $\kp<0$  the  dynamics is correctly defined for all the values of $r$, but  when $\kp>0$, the Hamiltonian $\wt{{\cal T}}_c$ has a singularity at $r =\kp$. Note that  $  \mu_c$ is only positive when $r>\lambda$, so in this case the   dynamics is only defined outside this circle,  that is, the region in which $\wt{{\cal T}}_c$  is positive definite.    
We note that  this PDM shows a certain similarity with the coefficient (related to the singularity) in the Schwarzschild metric.

The separability of the HJ  equation in cylindrical  coordinates $(r,\phi,z)$
$$
 \frac{r}{r - \kp} \left[      \left(\fracpd{W}{r}\right)^2 + \frac{1}{r^2}\left(\fracpd{W}{\phi}\right)^2 + 
   V_c(r,\phi) \left(\fracpd{W}{z}\right)^2    \right]   = 2  E \,, 
$$
leads to the following constants of motion
\begin{eqnarray}
 && \wt K_{c1}  =  p_z  \,,\qquad
 \wt  K_{c2}  = p_\phi^2  +  \Bigl(  \frac{k_2}{\sin^2\phi}  +  \frac{k_3 \cos\phi}{\sin^2\phi } \Bigr)\,p_z^2\,,\nonumber\\[2pt] 
&& \wt  K_{c3}  = r^2 p_r^2 +  k_1r \,p_z^2   -  2 r(r-\kp)   \wt {\cal T}_c\, . 
\label{ge}
\end{eqnarray}
Note that  $\wt  K_{c2} +  \wt  K_{c3} = 0$ and that  $\{ \wt  K_{c1}\,, \wt  K_{c2}\} = 0$. 
The separability of the HJ  equation in parabolic-cylindrical coordinates $(\pa,\pb,z)$,
$$
  \frac{1}{\pa^2+\pb^2 -2 \kp}\left[ \left(\fracpd{W}{\pa}\right)^2 + \left(\fracpd{W}{\pb}\right)^2+(\pa^2+\pb^2) V_c(\pa,\pb) \left(\fracpd{W}{z}\right)^2   \right] =  2 E \, ,
$$
also leads to three constants of motion but only one is  functionally independent with respect to the above ones, namely
\beq
\wt   J_{c2}  = p_\pa^2 + \left( k_1 + \frac{k_2-k_3}{\pa^2} \right) p_z^2+ 2 (\kp- \pa^2)\,\wt {\cal T}_c \, .
\label{gf}
\eeq

We summarize the results in the following proposition. 

\begin{proposition}  \label{proposition11}
The 3D $\kp$-dependent geodesic Hamiltonian  $\wt{\cal T}_c(\kp)$ (\ref{hax}) with ${{\cal T}}_c$   (\ref{ec}) and PDM  (\ref{mass3}) 
$$
  \wt{\cal T}_c = \frac 1{\mu_c} \, {\cal T}_c = \frac{1}{2\mu_c(x,y)}\left( p_x^2 + p_y^2 + V_c(x,y)\,p_z^2 \right) , {\qquad}  \mu_c(x,y) = 1 - \frac{\kp}{r}  \, , 
$$ 
 is HJ separable in cylindrical   $(r,\phi,z)$ and parabolic-cylindrical $(\pa,\pb,z)$ coordinates.  This   is endowed with four independent constants of motion: the Hamiltonian itself, $\wt{\cal T}_c$,  along with $\wt K_{c1}, \wt K_{c2}$ (\ref{ge}) and $\wt J_{c2}$ (\ref{gf}).
The three integrals $\wt {\cal T}_c, \wt K_{c1}, \wt K_{c2}$ are mutually in involution. 
\end{proposition}

Notice that the functions $\wt K_{c1}$ and $\wt K_{c2}$ are $\kp$-independent, so   coinciding with the original constants    (\ref{de}). 
Clearly, when taking  the  limit $\kp\to 0$ the results given in Proposition  3 are recovered.

\subsection{Geodesic Hamiltonian $\wt{{\cal T}}_d$  with PDM from Kepler--Coulomb II}

As far as the last family is concerned, we consider the geodesic Hamiltonian  $\wt{{\cal T}}_d$ (\ref{hax}) with ${{\cal T}}_d$   (\ref{ed}) and  with the same previous PDM $\mu_d\equiv \mu_c$ (\ref{mass3}).

The separability of the HJ equation in parabolic-cylindrical coordinates of type I $(\pa,\pb,z)$,
$$
  \frac{1}{\pa^2+\pb^2 - 2\kp}\left[ \left(\fracpd{W}{\pa}\right)^2 + \left(\fracpd{W}{\pb}\right)^2 
  +(\pa^2+\pb^2) V_d(\pa,\pb) \left(\fracpd{W}{z}\right)^2   \right] =  2 E \,, 
$$
yields  three first integrals 
\begin{eqnarray}
&&\wt  K_{d1}  =   p_z \,,  \qquad
\wt   K_{d2}  = p_\pa^2 + \left(   k_1 +  2 k_2 \pa \right) p_z^2 + 2 \left(\kp-\pa^2\right)\wt   {\cal T}_d  \,,\nonumber\\[2pt] 
&& \wt  K_{d3}  = p_\pb^2 + \left(   k_1 + 2 k_3 \pb \right) p_z^2 + 2 \left(\kp-\pb^2\right) \wt  {\cal T}_d \,,  
\label{gg}
\end{eqnarray}
but note that  $\wt K_{d2}+ \wt K_{d3} = 0$. Moreover, $\wt K_{d1}$ and $\wt K_{d2}$ are in involution, i.e.  $\{\wt K_{d1}\,,\wt K_{d2}\} = 0$. 
The separability of the HJ equation in parabolic-cylindrical coordinates of type II $(\al,\be,z)$,
$$
  \frac{1}{\al^2+\be^2 - 2\kp}\left[ \left(\fracpd{W}{\al}\right)^2 + \left(\fracpd{W}{\be}\right)^2 
  +(\al^2+\be^2) V_d(\al,\be) \left(\fracpd{W}{z}\right)^2   \right] =  2 E \,, 
$$
also gives rise to three constants of motion,  but only the following one  is actually functionally independent,
\beq
\wt J_{d2}  = p_\al^2 + \left(   k_1 + \sqrt{ 2} ( k_2+k_3)  \al \right) p_z^2 +2\left(\kp -   \al^2\right) \wt  {\cal T}_d    \, .  
\label{gh}
\eeq

In this way, we obtain the $\kp$-generalisation of Proposition 4 as follows.

\begin{proposition}  \label{proposition12}
The 3D $\kp$-dependent geodesic Hamiltonian  $\wt{\cal T}_d(\kp)$ (\ref{hax}) with ${{\cal T}}_d$   (\ref{ed}) and PDM (\ref{mass3}) 
$$
  \wt{\cal T}_d = \frac 1{\mu_d} \, {\cal T}_d = \frac{1}{2\mu_d(x,y)}\left( p_x^2 + p_y^2 + V_d(x,y)\,p_z^2 \right) , {\qquad}  \mu_d(x,y) = 1 - \frac{\kp}{r}  \, , 
$$ 
is HJ separable in two sets of parabolic-cylindrical  coordinates $(\pa,\pb,z)$ and   $(\al,\be,z)$ which are related by a rotation. 
This system is  endowed with four  independent  constants of motion:  the Hamiltonian  $\wt{\cal T}_d$, together with   
  $\wt K_{d1}, \wt K_{d2}$ (\ref{gg}) and $\wt J_{d2}$ (\ref{gh}).
The three first integrals $\wt{\cal T}_d, \wt K_{d1}, \wt K_{d2}$ are mutually in involution. 
\end{proposition} 

   The two constants   of motion $\wt K_{d2}$ and $\wt J_{d2}$ are the $\kp$-dependent version of   the generalised Laplace--Runge--Lenz 2-vector (\ref{xa}),  which   in the coordinates  $(\pa,\pb,z)$ read as 
\begin{eqnarray}
&& \wt  K_{d2}  =   \frac{\pb^2 p_\pa^2-\pa^2 p_\pb^2+\kp (p_\pb^2-p_\pa^2  ) }{\pa^2 + \pb^2-2\kp} + \left(
 \frac{k_1( \pb^2-\pa^2 ) + 2 k_2  ( \pb^2 -\kp )  \pa- 2 k_3    ( \pa^2-\kp )\pb  }{\pa^2 + \pb^2-2\kp} \right) p_z^2  \,, \nonumber\\[2pt] 
&&\wt    J_{d2}  =  \frac{(\pa  p_\pb -\pb  p_\pa )(\pa  p_\pa-\pb  p_\pb)-2\kp\, p_\pa p_\pb}{\pa^2 + \pb^2-2\kp}  \nonumber\\[2pt] 
&&\qquad\qquad  - 
\left( \frac{2 k_1 \pa \pb  + k_2 (\pa^2 - \pb^2+2\kp) \pb  - k_3  (\pa^2 - \pb^2-2\kp)  \pa}{\pa^2 + \pb^2-2\kp} \right) p_z^2 \, .
\label{xag}
\end{eqnarray}

 So far, we have  constructed 3D systems  from 2D ones by following the Eisenhart prescription. Next, a   generalisation of such a procedure has been achieved  by considering  a PDM  in the 2D system, so that we have obtained   PDM functions depending   only on the two first coordinates.  In this respect, we remark that one could also consider an alternative construction   inserting a PDM term in the 3D system (\ref{CaHa}) obtained once the Eisenhart lift has been applied.
This would allow  a further generalisation  such that  the new  metric $\dd \wt\sigma_{r}^2$ and   the new geodesic Hamiltonian $\wt{\cal T}_r$ would be    (see (\ref{ba}) and (\ref{CaHa}))
\bea
&&  \dd \wt\sigma_r^2 =\mu_r\,  \dd \sigma_r^2= \mu_r(x,y,z) \left( \dd x^2 +  \dd y^2 + \frac{\dd z^2}{V_r(x,y)} \right) , 
\nonumber\\[2pt]
&&
 \wt{\cal T}_r = \frac 1{\mu_r} \,   {\cal T}_r = \frac{1}{2\mu_r(x,y,z)}\left( p_x^2 + p_y^2 + V_r(x,y)\,p_z^2 \right) , \qquad r=a,b,c,d\, . 
\label{hax2}\nonumber
\eea
  However this point requires a deeper analysis as well as other related question when looking for alternative expressions to those here studied in (\ref{mass1}), (\ref{mass2}) and (\ref{mass3}).

\sect{3D   Hamiltonians $\widetilde{\cal H}$ with a position-dependent mass} \label{section7}

The last step in our approach is to add a potential $\wt \pot_r$ to the geodesic Hamiltonian with a PDM $\wt{\cal T}_r $ (\ref{hax}). 
We impose that the resulting Hamiltonian, $\wt{\cal H}_r =\wt{\cal T}_r + \wt \pot_r$, be once again separable,  so superintegrable, in the same two sets of coordinate systems corresponding to $\wt{\cal T}_r $ and fulfilling  the same three requirements assumed at the beginning of  Section 6. Therefore, we introduce the PDM $\mu_r$ in the  Hamiltonian (\ref{CaHamX}),  $\wt{\cal H}_r={\cal H}_r/\mu_r$,  and by taking into account (\ref{CaHamXb}) and (\ref{CaHam}), we are led to consider  the following family of Hamiltonians $(r=a,b,c,d)$
\bea
&&\wt{\cal H}_r= \wt{\cal T}_r + \wt \pot_r= \frac {1}{\mu_r}\, {\cal T}_r  +  \frac {1}{\mu_r}\, {\cal U}_r \nonumber\\[2pt]
&&\quad     = \frac{1}{2\mu_r(x,y)}\left( p_x^2 + p_y^2 + V_r(x,y)\,p_z^2 \right)+ \frac{1}{ \mu_r(x,y)} \left( U_r(x,y) +  V_r (x,y) Z(z) \right) ,
\label{7a}
\eea
where $U_r(x,y)$ is  the  function  to be determined for each family while  $ Z(z)$ is  an arbitrary smooth function.
Then, as in Section 5,  it is found that  $U_r$ keeps the same formally form as the 2D Euclidean potential $V_r$ but with different coefficients $t_i$ instead of $k_i$ $(i=1,2,3)$.

We display the resulting 3D superintegrable Hamiltonians (\ref{7a}) with a PDM  along with their independent constants of motion,   $\wt {\cal K}_{ri}$ and  $\wt {\cal J}_{ri}$,  in Table~\ref{Table1}. 
We stress that, in fact,  this comprises  the main results of the paper that generalise all the previous ones.

Now some remarks are in order.

\begin{itemize}
\item Notice that if we set $t_i=0$  and $Z(z)\equiv 0$ in Table~\ref{Table1}, then  $ \wt \pot_r\equiv 0$ and $ \wt{\cal H}_r\to  \wt{\cal T}_r$,  $\wt {\cal K}_{ri}\to \wt{ K}_{ri}  $,   $\wt{\cal J}_{ri}\to \wt{ J}_{ri}$ recovering Propositions 9--12 of Section 6 for the  geodesic Hamiltonian with a PDM $\wt{\cal T}_r$.

\item The limit $\la\to 0$, so $\mu_r\to 1$, in Table~\ref{Table1} gives rise to the limits 
 $ \wt{\cal H}_r\to   {\cal H}_r$,  $\wt {\cal K}_{ri}\to  {\cal K}_{ri}  $,   $\wt{\cal J}_{ri}\to  {\cal  J}_{ri}$, thus reproducing   the results presented in Propositions 5--8 of Section 5 for the  superintegrable   Hamiltonian $ {\cal H}_r$.

\item We remark that the PDM $\mu_a=1-\la r^2$ is just the same conformal factor used in the construction of the so called Darboux III oscillator~\cite{BurgosAnnPh11,PhysD2008}.  
Recall that this is   an exactly solvable model which can be regarded as a  generalisation of the isotropic oscillator to a conformally flat space of nonconstant curvature, being  a particular Bertrand space~\cite{Perlick92, Ballestetal08ClQGr, ComRunge}. 
Explicitly,  if we consider $\wt{\cal H}_a$ in  Table~\ref{Table1}  and set $k_2=k_3=t_i=0$ ($i=1,2,3$), $Z(z)\equiv 0$, $k_1=\omega^2$ and   $p_z=\sqrt{2}$,  we obtain   the following  2D Hamiltonian
$$
\wt{\cal H}_a\equiv \wt{\cal T}_a=\frac{1}{2(1-\la r^2)}\left(p_x^2+p_y^2+ \omega^2 r^2 \right)
= \frac{p_x^2+p_y^2}{2(1-\la r^2)}   + \frac{\omega^2 r^2}{2(1-\la r^2)}     \, ,
$$
which is the 2D counterpart of the Darboux III oscillator (change $\la\to -\la$ in~\cite{BurgosAnnPh11}). Consequently, the Hamiltonian $\wt{\cal H}_a$ in  Table~\ref{Table1} turns out to be  a 3D superintegrable generalisation of that system by breaking spherical symmetry.

\item The two KC families have the same PDM $\mu_c=\mu_d= 1 -  {\la}/{r}$. We stress that this is just the conformal factor  considered in the construction of an exactly solvable deformation of the KC problem~\cite{Bal14AnnPhys}; such a system is   related to a reduction \cite{IwaiKat94} of the geodesic motion on the Taub--NUT space which turns out to be another  Bertrand space~\cite{Perlick92, Ballestetal08ClQGr, ComRunge}. 
In particular, if we set     $k_2=k_3=t_i=0$ ($i=1,2,3$), $Z(z)\equiv 0$, $k_1=-k$ and   $p_z=\sqrt{2}$ in either  $\wt{\cal H}_c$  or $\wt{\cal H}_d$ in  Table~\ref{Table1}, we find that
$$
\wt{\cal H}_c\equiv \wt{\cal H}_d\equiv \wt{\cal T}_c\equiv \wt{\cal T}_d=\frac{r}{2(r-\la )}\left(p_x^2+p_y^2- \frac{2k}{r}   \right) =\frac{r(p_x^2+p_y^2)}{2(r-\la )}- \frac{k}{r-\la }  \,   ,  
$$
which is the 2D version of the deformed KC problem (change $\eta\to -\la$ in~\cite{Bal14AnnPhys})
Therefore, the Hamiltonians $\wt{\cal H}_c$  and $\wt{\cal H}_d$ in  Table~\ref{Table1} provide two different possible 3D superintegrable generalisations for the above system.
\end{itemize}


 \begin{table}[htbp]
 {\footnotesize{
\caption{\footnotesize The four families of 3D superintegrable    Hamiltonians $\widetilde{\cal H}_r$ ($r=a,b,c,d$) with a PDM. For each family, we write the PDM  $\mu_r$,   the 2D potential  $V_r$   ($U_r$ is the same but with coefficients $t_i$),   and the four  independent constants of motion  in the two sets of separable coordinates.  All of them share a common integral $\wt {\cal K}_{r1}= p_z^2 + 2 Z(z)$.}
\label{Table1}
 \begin{center}
\noindent
\begin{tabular}{l}
\hline
\\[-0.25cm] 
\quad Generic Hamiltonian:\quad$\displaystyle{ \wt{\cal H}_r    = \frac{1}{2\mu_r}\bigl( p_x^2 + p_y^2 + V_r \,p_z^2 \bigr)+ \frac{1}{ \mu_r } \bigl( U_r  +  V_r Z(z) \bigr) }$\\
\\[-0.25cm] 
\hline
\\[-0.2cm]
$\bullet$ Family a: From isotropic  oscillator \qquad PDM:  $\mu_a = 1 - \kp(x^2+y^2) = 1-\kp  r^2  $\\[2pt]  
\phantom{$\bullet$}   Separable in Cartesian $(x,y,z)$ and  cylindrical  $(r,\phi,z)$ coordinates\\ [0.1cm]
$\displaystyle{V_a=  \frac 12 k_1 (x^2 + y^2) + \frac{k_2}{x^2}  + \frac{k_3}{y^2}= \frac 12 k_1 r^2 + \frac{k_2}{r^2\cos^2\phi}  + \frac{k_3}{r^2\sin^2\phi} }$\\ [0.25cm]
$\displaystyle{ \wt{\cal K}_{a2} = p_x^2 +\left( \frac 12  k_1x^2+\frac{k_2}{x^2}\right)\left(p_z^2  + 2 Z(z)\right)+   t_1 x^2  +  \frac{2t_2}{x^2}  +  2 {\kp} x^2 \wt{{\cal H}}_a  }$\\ [0.3cm]
$\displaystyle{ \wt {\cal K}_{a3} = p_y^2 +\left( \frac 12   k_1y^2+\frac{k_3}{y^2}\right)\left(p_z^2  + 2 Z(z)\right)   + t_1 y^2  +  \frac{2t_3}{y^2}   +  2 {\kp} y^2 \wt{{\cal H}}_a}$\\ [0.3cm]
$\displaystyle{ \wt  {\cal J}_{a2}= p_\phi^2  +\left(   \frac{k_2}{\cos^2\phi}  +  \frac{k_3}{\sin^2\phi}\right)  \left(p_z^2  + 2 Z(z)\right) +    \frac{2t_2}{\cos^2\phi}  +  \frac{2t_3}{\sin^2\phi}  }$\\ [0.3cm]
Independent integrals: $\wt {\cal K}_{a1},\wt {\cal K}_{a2},\wt {\cal K}_{a3},\wt {\cal J}_{a2}$ with $ \wt{\cal H}_a  = \frac{1}{2}(\wt {\cal K}_{a2} +  \wt {\cal K}_{a3} )$\qquad
In involution:   $\wt {\cal K}_{a1},\wt {\cal K}_{a2},\wt {\cal K}_{a3}$
\\[0.2cm]
\hline
\\[-0.2cm]
$\bullet$ Family b: From anisotropic  oscillator\qquad PDM:   $\mu_b =1 - \kp\,  x =1-\frac 12\kp(\pa^2-\pb^2)  $\\[2pt] 
\phantom{$\bullet$}   Separable in Cartesian $(x,y,z)$ and  parabolic-cylindrical $(\pa,\pb,z)$  coordinates\\ [0.1cm]
$\displaystyle{V_b= \frac 12  k_1 (4 x^2 + y^2) + \frac{k_2}{y^2}  + k_3 x    =  \frac{1}{\pa^2+\pb^2}\left[\frac {k_1}2(\pa^6+\pb^6)  +  k_2 \left(\frac{1}{\pa^2}+\frac{1}{\pb^2}\right)  + \frac {k_3}2 (\pa^4-\pb^4)  \right]  }$\\ [0.3cm]
$\displaystyle{ \wt{\cal K}_{b2} =  p_x^2  + \left(2 k_1x^2+k_3 x\right) \bigl(p_z^2  +  2 Z(z)\bigr)  +  4 t_1 x^2  + 2 t_3 x  + 2 {\kp} x   \wt{{\cal H}}_b  }$\\ [0.2cm]
$\displaystyle{ \wt {\cal K}_{b3} =   p_y^2  + \left( \frac 12 k_1y^2 + \frac{k_2}{y^2}\right)\bigl(p_z^2  +  2 Z(z)\bigr) +   t_1 y^2  +  \frac{2 t_2}{y^2}}$\\ [0.3cm]
$\displaystyle{ \wt  {\cal J}_{b2}= p_\pa^2 + \left( \frac{k_1}2  \pa^6+ \frac{k_2}{\pa^2} + \frac{k_3}2  \pa^4\right)\bigl(p_z^2  +  2 Z(z)\bigr) +  { t_1} \pa^6+ \frac{2t_2}{\pa^2} + { t_3} \pa^4   -   \pa^2\left( 2-  \kp    \pa^2\right) 
\wt{{\cal H}}_b }$\\ [0.3cm]
Independent integrals: $\wt {\cal K}_{b1},\wt {\cal K}_{b2},\wt {\cal K}_{b3},\wt {\cal J}_{b2}$ with $ \wt{\cal H}_b  = \frac{1}{2}(\wt {\cal K}_{b2} +  \wt {\cal K}_{b3} )$\qquad
In involution:   $\wt {\cal K}_{b1},\wt {\cal K}_{b2},\wt {\cal K}_{b3}$
\\[0.2cm]
\hline
\\[-0.2cm]
$\bullet$ Family c: From Kepler--Coulomb I \qquad PDM:  $  \mu_c= 1 -  {\kp}/{r}= 1-2\kp/(\pa^2+\pb^2)$ \\[2pt] 
\phantom{$\bullet$}   Separable in  cylindrical  $(r,\phi,z)$  and  parabolic-cylindrical $(\pa,\pb,z)$  coordinates\\ [0.1cm]
$\displaystyle{V_c= \frac{k_1}{r}  +  \frac{k_2}{r^2\sin^2\phi}  +  \frac{k_3 \cos\phi}{r^2\sin^2\phi}  =   \frac{1}{\pa^2+\pb^2}\left[2 k_1  +  k_2 \left(\frac{1}{\pa^2}+\frac{1}{\pb^2}\right)  
 +k_3 \left(\frac{1}{\pb^2}-\frac{1}{\pa^2}\right)   \right]  }$\\ [0.3cm]
$\displaystyle{ \wt{\cal K}_{c2} =   p_\phi^2 +\left(   \frac{k_2}{\sin^2\phi}  +  \frac{k_3\cos\phi}{\sin^2\phi}\right)  \bigl(p_z^2  + 2 Z(z)\bigr)  +  \frac{2t_2}{\sin^2\phi}  +  \frac{2t_3\cos\phi}{\sin^2\phi}    }$\\ [0.3cm]
$\displaystyle{ \wt  {\cal J}_{c2}=  p_\pa^2 + \Bigr(  k_1 + \frac{k_2-k_3}{\pa^2} \Bigl) \bigl(p_z^2  + 2 Z(z)\bigr)  +   2t_1 + \frac{2(t_2-t_3)}{\pa^2} + 2 (\kp-\pa^2 )\wt{\CaH}_c }$\\ [0.3cm]
Independent integrals: $\wt {\cal H}_{c},\wt {\cal K}_{c1},\wt {\cal K}_{c2},\wt {\cal J}_{c2}$ \qquad
In involution:   $\wt {\cal H}_{c}, \wt {\cal K}_{c1},\wt {\cal K}_{c2}$
\\[0.2cm]
\hline
\\[-0.2cm]
$\bullet$ Family d: From Kepler--Coulomb II \qquad PDM:  $  \mu_d= 1-2\kp/(\pa^2+\pb^2)= 1-2\kp/(\al^2+\be^2)$ \\[2pt] 
\phantom{$\bullet$}   Separable in parabolic-cylindrical    $(\pa,\pb,z)$ and    $(\al,\be,z)$  coordinates\\ [0.1cm]
$\displaystyle{V_d=2 \frac{k_1  +  k_2 \pa + k_3 \pb}{\pa^2+\pb^2}  =  \frac{2 k_1  +  k_2 \sqrt{2}(\al+\be)  + k_3 \sqrt{2}(\al-\be)}{\al^2+\be^2}}$\\ [0.2cm]
$\displaystyle{ \wt{\cal K}_{d2} =   p_\pa^2 + \left(   k_1 +  2 k_2 \pa \right) \bigl(p_z^2  + 2 Z(z)\bigr) +  2t_1 +  4 t_2 \pa  + 2 \left(\kp-\pa^2\right)\wt   {\cal H}_d  }$\\ [0.2cm]
$\displaystyle{ \wt  {\cal J}_{d2}= p_\al^2 + \bigl(   k_1 + \sqrt{ 2} ( k_2+k_3)  \al \bigr) \bigl(p_z^2  + 2 Z(z)\bigr) +2t_1+2\sqrt{2}(t_2+t_3)\al  +2\left(\kp -   \al^2\right) \wt  {\cal H}_d }$\\ [0.2cm]
Independent integrals: $\wt {\cal H}_{d},\wt {\cal K}_{d1},\wt {\cal K}_{d2},\wt {\cal J}_{d2}$ \qquad
In involution:   $\wt {\cal H}_{d}, \wt {\cal K}_{d1},\wt {\cal K}_{d2}$
\\[0.25cm]
\hline
\end{tabular}
\end{center}
}}
 \end{table}


\newpage

\section{Concluding remarks and outlook}

As already  observed in the Introduction, the first studies on superintegrability were mainly concerned with the analysis of potentials defined on (2D  and 3D) Euclidean spaces. 
Then, a second step was the study of potentials on Riemannian  spaces of   constant curvature (spherical and hyperbolic geometries), and only recently the  existence of superintegrable systems on more general Riemannian spaces has become a matter of study. 
In this last situation the problem  becomes much more complicated since the superintegrability  depends, not only on the potential,  but also on the coefficients of the non-Euclidean metric. 
In particular, the existence of integrals of motion for the free particle (geodesic motion determined by the metric) must be studied; only when this question has been solved, the existence of potentials with superintegrability can be analysed.  

We have here applied the Eisenhart formalism to the four families of 2D superintegrable Euclidean Hamiltonians $H_r=T+V_r$, $(r=a,b,c,d)$, and we have studied the separability of the four 3D geodesic Hamiltonians 
$$
 {\cal T}_r = \frac{1}{2} \Bigl( p_x^2 + p_y^2 + V_r(x,y)\,p_z^2 \Bigr)
   \,, 
$$  
and then we have extended the study to the separability of Hamiltonians with the addition of a potential 
$$
 {\CaH}_r=  {\cal T}_r+\pot_r = \frac{1}{2} \Bigl( p_x^2 + p_y^2 + V_r(x,y)  p_z^2 \Bigr) + \Bigl(U_r(x,y) +  V_r(x,y) Z(z) \Bigr)
\, . 
$$
Furthermore, we have also study the separability of the 3D geodesic Hamiltonians with a PDM  $\mu_r$
$$
 \wt {\cal T}_r(\kp) = \frac{1}{\mu_r}  {\cal T}_r= \frac{1}{2\mu_r(x,y)} \Bigl( p_x^2 + p_y^2 + V_r(x,y) p_z^2\Bigr) 
  \,, 
  $$
and finally, as our most general result  shown in Table 1, we have   obtained superintegrable    Hamiltonians with both potential and PDM:
$$
 \wt {\CaH}_r(\la) =   \frac{1}{\mu_r}  {\cal H}_r=  \frac{1}{2\mu_r(x,y)} \Bigl( p_x^2 + p_y^2 + V_r(x,y) p_z^2\Bigr) 
 + \frac{1}{\mu_r(x,y)} \Bigl(U_r(x,y) +  V_r(x,y) Z(z) \Bigr)   \, . 
$$
We remark that  the PDM $\mu_r$ depends on a parameter $\la$ in such a way that  the superintegrability is preserved for all the values of $\la$   and that the Hamiltonian  $ \wt {\CaH}_r(\la)$ can be considered as continuos deformation  of the previously studied Hamiltonian  $ {\CaH}_r$. 

  We conclude with the following open problems. 
First, all the constants of motion we have obtained are a straightforward consequence of the existence of symmetries (in most of cases hidden symmetries); so it would be convenient to study the properties of these symmetries from a geometric approach  (that is, symplectic formalism and Lie algebra of vector fields). 
It would also be convenient the study of higher-order constants of motion; this means to study the existence of Killing tensors ${\bf K}$ of valence   $p>2$ (see Section \ref{section55}). In this respect, the results of this paper can be regarded as a first step in this direction; in fact, the obtention of superintegrable systems with higher-order  integrals and with broken spherically symmetry   remains as a non-trivial  open task.
Second, we have applied the geometric Eisenhart formalism starting with superintegrable Hamiltonians $H_r$ defined on the Euclidean plane; a possible generalisation should  be to consider as starting point superintegrable systems defined not on the Euclidean plane but on the  2D spaces with constant curvature, that is,  either on the sphere (positive curvature) or  on the hyperbolic plane (negative curvature). 
Third, it is known that HJ separability is related to separability of the Schr\"odinger equation; hence  it would be also convenient to study the quantum counterparts  of the Hamiltonian systems studied throughout  this paper. 
The  Eisenhart lift has been related to the properties of Killing tensors defined on the Riemannian space; thus this is also a matter to be studied. 

Finally, we recall the comment a the end of Section \ref{section6}; the PDM $\mu_r$, $r=a,b,c,d$, of the systems we have studied  are functions of the initial varia\-bles $x$ and $y$  and independent of the new degree of freedom; this two-dimensional dependence of the PDMs  is not a limitation of the approach but a property directly related with the particular form of the Eisenhart formalism considered in this paper.  Nevertheless, a  possible extension of these systems can be obtained by considering the more general case of $z$-dependent PDM; we have already obtained some results (introducing an additional term coupling the three coordinates and depending of a second parameter) but this point remains as a question for future work.  

\section*{Appendix: A geometric approach to Eisenhart lift}
\setcounter{equation}{0}
\renewcommand{\theequation}{A.\arabic{equation}}

Natural Lagrangians, also called Lagrangians of mechanical type, are defined by a differentiable function $V$ in a (pseudo-)Riemannian manifold $(M,g)$. We denote $L_{g,V}$ such Lagrangians:
\begin{equation}
  L_{g,V}(q,v)=\frac 12 \,g_q(v,v)-(\tau_M^*V)(q,v)=\frac 12 \,g_q(v,v)-V(q),  \label{mechL}
\end{equation}
where $\tau_M:TM\to M$ is the tangent bundle projection,  i.e. the Lagrangian function is of the form $L_{g,V}=T_g-\tau_M^* V$, where the function $T_g\in C^\infty (TM)$  represents the kinetic  energy. 
Here we follow the notation of \cite{CGMS14} where more mathematical details can be found. In local coordinates $(q^i)$ in an open
 set $U$ of $M$ and 
the associated coordinates $(q^i,v^i)$ in its tangent bundle, the local expressions for the Riemannian metric $g$ and kinetic energy  $T_g$ are respectively written as 
$$
 g=g_{ij}(q)\,\dd q^i\otimes \dd q^j,\qquad 
 T_g(v)=\frac 12\, g_{ij}(\tau_M(v))\,v^iv^j.\label{kinenerg}
$$

Nondegeneracy of the Riemann structure means that $L_{g,V}$ is a regular Lagrangian and defines a Hamiltonian dynamical system
 $(TM,\omega_{L_{g,V}}, E_{L_{g,V}})$ (see \cite{CGMS14} and references therein)
and the dynamical vector field $\Gamma_{L_{g,V}}$, Hamiltonian vector field defined by the energy function $E_L$, i.e. defined by
$i(\Gamma_{L_{g,V}})\omega_{L_{g,V}}=\dd E_{L_{g,V}}$, takes the form 
$$
\Gamma_{L_{g,V}}(q,v)= v^i\fracpd{}{q^i}- \left( \Gamma_{jk}^i(q)v^jv^k+ g^{ij}(q)\frac{\partial V}{\partial q^j}(q)\right)\fracpd{}{v^i}, \label{dynvfV}
$$
where  $\Gamma_{jk}^i$ are the Christoffel symbols of the second kind with respect to 
 the Levi--Civita connection defined by the metric $g$, given by 
$$
\Gamma^i_{jk}=\frac 12  g^{il} \left(\fracpd{g_{lj}}{q^k}+\fracpd{g_{lk}}{q^j}-\fracpd{g_{jk}}{q^l}\right).\label{Gijk}
$$
Consequently, the curves in the manifold $M$ whose tangent lifts are integral curves of 
 $\Gamma_{T_g}$ are such that $\nabla_{\dot \gamma}{\dot \gamma}+\widehat g^{-1}(\dd V)=0$, with local coordinate expression 
\begin{equation}
g_{li}\left(\ddot q^i+\Gamma^i_{jk}\,\dot q^j\,\dot q^k\right)=-\fracpd V{q^l} \,, \qquad l=1,\ldots,\dim M.\label{dyneq}
\end{equation}

In the particular case $V=0$ we see that such curves in $M$ are but the geodesics of the Riemannian metric and the geodesic motion is called free motion. 
Recall that the arc-length of a curve $\gamma$  in $M$ between the points $\gamma(t_1)$ and $\gamma(t_2)$ is given by 
$$
\int_{t_1}^{t_2} \sqrt{g(\dot\gamma,\dot\gamma)}\ \dd t \,,
$$
and the extremal length curves are those of  the action defined by the Lagrangian $\ell (v)= \sqrt{g(v,v)} $, even if we have to restrict ourselves to 
the open submanifold $T_0M=\{ v\in TM\mid v\ne 0\}$ in order to preserve the differentiability.  Then we can consider the Lagrangian
$\ell(v)=\sqrt{2\,{T_g}(v)}$, which  is a singular Lagrangian whose relation to the Lagrangian $T_g=L_{g,0}$ has been studied in \cite{CaLop91}.

Note that since 
   $$
  \fracpd{\ell}{\dot q^i}= \frac{g_{ij}\dot q^j} {\ell}, \qquad   \fracpd{\ell}{q^i}= \frac 1{2\ell}\left(\fracpd {g_{jk}}{q^i}\dot q^j\,\dot q^k\right) ,
   $$
   the Euler--Lagrange equations of the Lagrangian $\ell$ are given by
   $$
   \frac \dd{\dd t}\left(\frac{g_{ij}\dot q^j} {\ell}\right)= \frac 1{2\ell}\left(\fracpd {g_{jk}}{q^i}\dot q^j\,\dot q^k\right).
   $$
   If we parametrize the curves by the arc-length, and then as $\dd s/\dd t=\ell$, $ \dot q^i=\ell\, q^{\prime i} =\ell \,\dd q^i/\dd s$, we find that    the previous system becomes 
   $$
   \frac \dd{\dd s}\left(g_{ij}\,q^{\prime j}\right)=\fracpd {g_{jk}}{q^i}\,q^{\prime j}\,q^{\prime k},
   $$
   which are the same equations as those obtained from $T_g$ with $q^{\prime k}$ instead of $\dot q^k$, and therefore in terms of such parametrization the 
   extremal arc-length curves  are geodesics of the corresponding metric. 
   
As $L_{g,V}$ is regular, we can carry out the Legendre transformation and then the elements  $(q,v)\in TQ$ 
correspond to elements  $(q,p)\in T^*Q$ in such a way that 
$$
p_i =\fracpd {L_{g,V}}{v^i}=\fracpd {T_g}{v^i}=g_{ik} (q)\, v^k\ \Longleftrightarrow\  v^i=g^{ij}(q)\, p_j\, ,
$$
with $g^{ik}(q)\, g_{kj}(q)=\delta^i_j$, and the Hamiltonian is nothing but the expression of the kinetic energy in terms of momenta plus the potential term
\begin{equation}
H=\frac 12\, g\bigl(\widehat g^{-1}(p), \widehat g^{-1}(p)\bigr)+V(q)=\frac 12\, g^{ij}\, p_i\, p_j+V(q) \,,\label{HgV}
\end{equation}
where $\widehat g:TM\to T^*M$ is the bundle map over the identity from the tangent bundle $\tau_M:TM\to  M$ 
 to the cotangent bundle $\pi_M:T^*M\to M$,  defined by $\langle\widehat g(v),w\rangle=g(v,w)$.

  In a seminal paper \cite{Eisenhart28} Eisenhart showed the possibility of relating the dynamical trajectories of 
a Lagrangian system of mechanical type (\ref{mechL}) with the projections on $M$ of extremal length curves on a extended manifold $\bar M=\mathbb{R}\times M$ 
with a Riemann structure 
$$
\bar g=\textrm{pr}_2^*g- \frac1{2\, V}\, \dd u\otimes \dd u \, ,\qquad u\in \mathbb{R} \, .
$$ 
More explicitly, if we assume that we choose $g_{00}$ as a function $A$ of the coordinates $q^1,\ldots,q^n$,  the arc-length reads
$$
\dd s^2=g_{ij}(q)\,\dd q^i\otimes \dd q^j+A(q)\,\dd u\otimes \dd u\, ,\label{dsEisen}
$$
with associated free motion described by
\begin{equation}
T_g=\frac 12\left(g_{ij}(q)\,v^i\, v^j+A(q)\,v_u^2\right).\label{TEisen}
\end{equation}
Then 
the equations of motion in terms of the arc-length $s$ turn out to be
$$
q^{\prime\prime i}+\Gamma^i_{jk}(q) \,q^{\prime j}\,q^{\prime k}-\frac 12 g^{ij} \fracpd A{q^j}\left(\frac{\dd u}{\dd s}\right)^2=0\, , \qquad i=1,\ldots,n,\label{eqEisen}
$$
together with the constant of motion corresponding to the invariance under translations in the variable $u$:
\begin{equation}
A(q)\, \frac{\dd u}{\dd s}=a\in\mathbb{R}\, .\label{pu1}
\end{equation}

For each value of the parameter $a$ we can use a new parameter $t$ such that $t=a\, s$ and then the differential equations reduce respectively to 
$$
 \ddot q^{ i}+\Gamma^i_{jk}(q) \,\dot q^{ j}\,\dot q^{ k}- g^{ij} \frac 1{2A^2}  \fracpd A{q^j}=0\, , \quad i=1,\ldots,n,\qquad A(q)\, \frac{\dd u}{\dd t}=1 \,.$$

Note that when $a=1$ the parameter $t$ coincides with $s$ and condition (\ref{pu1}) corresponds to set   $p_u=1$.

Suppose now a natural mechanical system in which the potential function $V$ is bounded
from below and that using the ambiguity in the choice of the potential we can assume that $V(q)>0$. 
Then if we choose $A=\dfrac 1{2V}$, the preceding 
system of differential equations  becomes equivalent to (\ref{dyneq})
$$
 \ddot q^{ i}+\Gamma^i_{jk}(q) \,\dot q^{ j}\,\dot q^{ k}+ g^{ij}   \fracpd V{q^j}=0 \, , \qquad i=1,\ldots,n \, .
$$
The free particle determined by the metric $\bar g$ is defined by the kinetic energy (\ref{TEisen})
and the Legendre transformation leads to the new Hamiltonian \cite{CarigliaAlv15}
\begin{equation}
  \bar H(q,u,p,p_u)=\frac 12\left(g^{ij}\,p_ip_j+V \,p_u^2\right),\label{geodH}
\end{equation}
which coincides for $p_u=\sqrt 2$ with (\ref{HgV}).

As pointed out by Benenti \cite{Ben97} the HJ separability of the Hamiltonian (\ref{HgV}) can be studied from the integrability of the geodesic Hamiltonian (\ref{geodH})

\section*{Acknowledgments}

JFC and MFR acknowledge support from research projects MTM2015-64166-C2-1 (MINECO, Madrid)  and DGA-E24/1 (DGA, Zaragoza).  FJH acknowledges   support by the  MINECO under project MTM2013-43820-P  and by the Spanish Junta de Castilla y Le\'on  under grant BU278U14 and VA057U16, as well as the warm hospitality at the  Department of
Theoretical Physics, University of Zaragoza, Spain.

  {\small
 }


\begin{thebibliography}{99}

\bibitem{Demkov}  Yu. N. Demkov, 
{\rm ``Symmetry group of the isotropic oscillator"}, Soviet Phys. JETP {\bf 36}, no.~9, 63--66, (1959).

\bibitem{Fradkin} D.M.  Fradkin, 
{\rm ``Three-dimensional isotropic harmonic oscillator and $SU_3$"}, 
Amer. J. Phys. {\bf 33}, 207--211 (1965).

\bibitem{FrMaSmUW65}  T.I. Fris ,   V. Mandrosov, Y.A. Smorodinsky,  M. Uhlir, and P. Winternitz,
{\rm ``On higher symmetries in quantum mechanics"},
Phys. Lett.  {\bf 16},  354--356  (1965).
 
\bibitem{Ev90Pra}  N.W.  Evans,  
{\rm ``Superintegrability in classical mechanics"},
Phys. Rev.  A  {\bf  41},  no. 10,  5666--5676  (1990).

\bibitem{GrPoSi95a}  C. Grosche,  G.S. Pogosyan,  and A.N.  Sissakian, 
{\rm ``Path integral discussion for Smoro\-dinsky--Winternitz potentials. 
I two-- and three-- dimensional Euclidean spaces"},
Fortschr. Phys.  {\bf  43},  no. 6,  453--521  (1995).


\bibitem{KaWiMiPo99}   E.G. Kalnins,  G.C. Williams,  W. Miller, and  G.S.  Pogosyan,
{\rm ``Superintegrability in the three--dimensional Euclidean space"},
J. Math. Phys.  {\bf  40},  no. 2,  708--725  (1999).

\bibitem{Ra97Jmp}  M.F.  Ra\~nada,  
{\rm ``Superintegrable $n=2$ systems, quadratic constants and potentials of Drach"}, 
J. Math. Phys.  {\bf  38},  no. 8,  4165--4178  (1997).

\bibitem{Tsi00Jpa}   A.V. Tsiganov ,
{\rm ``The Drach superintegrable systems"}, 
J. Phys. A: Math. Gen.   {\bf  33}, no. 41, 7407--7422  (2000).

\bibitem{RaSa01PLa}   M.F. Ra\~nada and  M. Santander, 
{\rm ``Complex euclidean super-integrable potentials, potentials of Drach,  and potential of Holt"},
Phys. Lett. A   {\bf  278},  271--279  (2001). 

\bibitem{Camp14}  R. Campoamor-Stursberg, 
{\rm ``Superposition of super-integrable pseudo-Euclidean  potentials in N = 2 with a fundamental constant of motion of arbitrary order in the momenta"}, 
J. Math. Phys.  {\bf  55},  042904  (2014).

\bibitem{GrPoSi95b}  C. Grosche,  G.S. Pogosyan,  and A.N.  Sissakian, 
{\rm ``Path integral discussion for Smoro\-dinsky--Winternitz potentials. 
II two-- and three-- dimensional sphere"}, 
Fortschr. Phys.  {\bf  43},  no. 6,  523--563  (1995).

\bibitem{RaSa99}  M.F. Ra\~nada and  M. Santander, 
{\rm ``Superintegrable systems on the two-dimensional sphere $S^2$ and the 
hyperbolic plane $H^2$"}, 
J. Math. Phys.  {\bf  40},  no. 10,  5026--5057  (1999).

\bibitem{KaKrPoMi01} E.G.  Kalnins,   J.M. Kress,    G.S. Pogosyan, and M.  Miller,  
{\rm ``Completeness of superintegrability in two-dimensional constant-curvature spaces"}, 
J. Phys. A: Math. Gen.   {\bf  34}, no. 22,  4705--4720  (2001).

\bibitem{KaKrWint02}   E.G.  Kalnins,   J.M. Kress,  and P.  Winternitz,  
{\rm ``Superintegrability in a two-dimensional space of nonconstant curvature"}, 
J. Math. Phys. {\bf  43},  no. 2,  970--983  (2002).


\bibitem{BaHeSantS03}  A. Ballesteros,   F.J. Herranz,  M. Santander, and  T. Sanz-Gil,
{\rm ``Maximal superintegrability on $N$-dimensional curved spaces"}, 
J. Phys. A: Math. Gen.   {\bf  36},  no. 7,   L93--L99  (2003).  

\bibitem{CRS07JPa}  J.F. Cari\~nena,    M.F. Ra\~nada, and  M. Santander, 
{\rm ``Superintegrability on curved spaces, orbits and momentum hodographs: revisiting a classical result by Hamilton"},   J. Phys. A: Math. Theor. {\bf 40}, no. 45, 13645--13666   (2007). 

\bibitem{BaHeMu13}   A.  Ballesteros,   F.J.  Herranz, and F. Musso,  
{\rm ``The anisotropic oscillator on the 2D sphere and the hyperbolic plane"}, 
Nonlinearity   {\bf  26},  no. 4,  971--990  (2013).  

\bibitem{BaBlHeMu14}   A. Ballesteros,  A.  Blasco,  F.J.  Herranz, and F. Musso,  
{\rm ``A new integrable anisotropic oscillator on the two-dimensional sphere and the hyperbolic plane"},  
J. Phys. A: Math. Theor.  {\bf 47},  no. 34,   345204  (2014).   

\bibitem{Ra14JPaTTWk}  M.F. Ra\~nada, 
{\rm ``The Tremblay-Turbiner-Winternitz  system on spherical and hyperbolic spaces : Superintegrability, curvature-dependent formalism and complex factorization"}, 
J. Phys. A: Math. Theor.  {\bf 47}, 165203  (2014).   

\bibitem{GonKas14AnnPhys}   C.  Gonera and M.  Kaszubska,    
{\rm Superintegrable systems on spaces of constant curvature"},  
Ann. Phys.  {\bf  364},  91--102  (2014).   

\bibitem{Mfran15PLa}  M.F. Ra\~nada, 
{\rm ``The Post-Winternitz system on spherical and hyperbolic spaces: a proof of the superintegrability making use of complex functions and a curvature-dependent formalism"},    
Phys. Lett. A {\bf 379}, no. 38, 2267--2271   (2015).   
 
\bibitem{MiPWJPa13}   W. Miller,  S.  Post,  and  P.  Winternitz,  
{\rm ``Classical and quantum superintegrability with applications"},    
J. Phys. A: Math. Theor.  {\bf  46},  423001  (2013).     

\bibitem{Eisenhart28}  L.P.  Eisenhart, 
{\rm ``Dynamical trajectories and geodesics"},     
Annals. Math. {\bf  30},  no. 1--4, 591--606 (1928--1929).  
[http://www.jstor.org/stable/1968307].  

\bibitem{Szydlowski98}  M. Szydlowski, 
{\rm ``The Eisenhart geometry as an alternative description of dynamics in terms of geodesics"},      
Gen. Relativity Gravitation {\bf  30}, no. 6, 887--914  (1998). 


\bibitem{SzydlowMacGu98}   M. Szydlowski,  A.J. Maciejewski,  and  J. Guzik,  
{\rm ``Dynamical Trajectories of Simple Mechanical Systems as Geodesics in Space with an Extra Dimension"},   
Internat. J. Theor.  Phys.{\bf  37}, no. 5, 1569  (1998). 

\bibitem{Benn06}   I.M.  Benn, 
{\rm ``Geodesics and Killing tensors in mechanics"},   
J. Math. Phys.  {\bf  47},  022903  (2006). 

\bibitem{Minguzzi07}  E.  Minguzzi, 
{\rm ``Eisenhart's theorem and the causal simplicity of Eisenhart's spacetime"}, 
Class. Quantum Grav. {\bf  24},  2781--2807  (2007). 

\bibitem{GibbonsHouri11}    G.W. Gibbons,  T. Houri,  D. Kubiznak,  and  C.M.  Warnick, 
{\rm ``Some spacetimes with higher rank Killing-Stackel tensors"}, 
Phys. Lett.  B   {\bf  700},  no. 1,   68--74  (2011). 

\bibitem{GalajPRd12}   A. Galajinsky, 
{\rm ``Higher rank Killing tensors and Calogero model"}, 
Phys. Rev. D  {\bf  85},   085002  (2012).

\bibitem{CarigGibbJmp14}  M. Cariglia   and  G. Gibbons, 
{\rm ``Generalised Eisenhart lift to the Toda chain"},   
 J. Math. Phys. {\bf  55},  no. 2, 022701  (2014).
 
\bibitem{CarigliaGetalJmp14}  M. Cariglia,  G.W. Gibbons,  J.W.  van Holten,  P.A.  Horvathy,  and P.M.  Zhang, 
{\rm ``Conformal Killing tensors and covariant Hamiltonian dynamics"},   
 J. Math. Phys. {\bf  55},  122702 (2014). 

\bibitem{Cariglia14Rmp}  M. Cariglia,  
``Hidden symmetries of dynamics in classical and quantum physics",
Rev. Mod. Phys.  86, 1283--1333 (2014). 

\bibitem{FilyukovGala15}  S.  Filyukov  and A.  Galajinsky,
{\rm ``Self-dual metrics with maximally superintegrable geodesic flows"}, 
Phys. Rev.  D    {\bf  91}, no. 10,  104020  (2015). 

\bibitem{CarigliaGala15}   M. Cariglia  and A.  Galajinsky, 
{\rm ``Ricci-flat spacetimes admitting higher rank Killing tensors"},  
Phys. Lett. B    {\bf  744},  320  (2015).

\bibitem{CarigliaAlv15}  M. Cariglia  and   F.K.  Alves,
{\rm ``The Eisenhart lift: a didactical introduction of modern geometrical concepts from Hamiltonian dynamics"},  
European J. Phys. {\bf  36},  no. 2,    025018   (2015). 

\bibitem{Vak05}    I.O.  Vakarchuk, 
{\rm ``The Kepler problem in Dirac theory for a particle with position-dependent mass"}, 
J. Phys. A: Math. Gen.  {\bf 38},  4727--4734  (2005).

\bibitem{RoRo05}  B. Roy  and  P. Roy,  
{\rm ``Effective mass Schr\"odinger equation and nonlinear algebras"}, 
Phys. Lett. A  {\bf 340},  70--73  (2005).

\bibitem{JiYiJ05} L. Jiang,   L.Z.  Yi ,  and    C.S.  Jia, 
{\rm ``Exact solutions of the Schr\"odinger equation with position-dependent mass for some Hermitian and non-Hermitian potentials"}, 
Phys. Lett. A  {\bf 345},   279--286  (2005).

\bibitem{Quesne06}  Ch.  Quesne, 
{\rm ``First-order intertwining operators and position-dependent mass Schr\"odinger equations in d dimensions"},  
Ann. Phys. {\bf 321},  no. 5  1221--1239 (2006).

\bibitem{CrNN07}  S.  Cruz y Cruz,  J.  Negro,  and  L.  Nieto, 
{\rm ``Classical and quantum position-dependent mass harmonic oscillators"}, 
Phys. Lett. A   {\bf 369},   400--406  (2007). 

\bibitem{Quesne07}  Ch.  Quesne, 
{\rm ``Spectrum generating algebras for position-dependent mass oscillator Schr\"odinger equations"}, 
J. Phys. A: Math. Theor.  {\bf  40}, 13107--13119 (2007). 

\bibitem{CrR09}  S.  Cruz y Cruz  and  O. Rosas-Ortiz, 
{\rm ``Position-dependent mass oscillators and coherent states"},  
J. Phys. A: Math. Theor.  {\bf   42},   185205 (2009). 

\bibitem{Yes10}  O.  Yesiltas,  
{\rm ``The quantum effective mass Hamilton-Jacobi problem"},  
J. Phys. A: Math. Theor.  {\bf  43},  095305  (2010).  

\bibitem{CobSch11}   H.  Cobian  and  A.  Schulze-Halberg, 
{\rm ``Time-dependent Schr\"odinger equations with effective mass in (2+1) dimensions: intertwining relations and Darboux operators"}, 
J. Phys. A: Math. Theor.  {\bf 44},   285301   (2011).  

\bibitem{BurgosAnnPh11}  A. Ballesteros,  A.  Enciso,   F.J.  Herranz, O.  Ragnisco,  and  O. Riglioni, 
{\rm ``Quantum mechanics on spaces of nonconstant curvature: the oscillator problem and superintegrability"}, 
 Ann. Phys.  {\bf  326},  no. 8,   2053--2073  (2011).

\bibitem{LimaVF12}   J.R.  Lima,  M. Vieira,  C. Furtado,  F. Moraes,  and  C. Filgueiras,  
{\rm ``Yet another position-dependent mass quantum model"}, 
J. Math. Phys. {\bf  53},   072101   (2012).  

\bibitem{Ran14Jmp}  M.F.  Ra\~nada,  
{\rm ``A quantum quasi-harmonic nonlinear oscillator with an isotonic term"},   
J. Math. Phys. {\bf  55}, 082108  (2014).  

\bibitem{GhoshRoy15}   D. Ghosh  and  B. Roy,  
{\rm ``Nonlinear dynamics of classical counterpart of the generalised quantum nonlinear oscillator driven by position-dependent mass"},  
Ann. Phys. {\bf  353},  222--237  (2015).    

\bibitem{MustJpa15}  O.  Mustafa,  
{\rm ``Position-dependent mass Lagrangians: nonlocal transformations, Euler-Lagrange invariance and exact solvability"},   
J. Phys. A: Math. Theor.  {\bf 48}, no. 22, 225206 (2015).    

\bibitem{Quesne15Jmp}    Ch. Quesne,  
{\rm ``Generalised nonlinear oscillators with quasi-harmonic behaviour:  Classical  solutions"}, 
J. Math. Phys. {\bf  56},  012903 (2015). 


\bibitem{Perlick92} V.  Perlick,  {\rm ``Bertrand spacetimes"},  
Class. Quantum Grav. {\bf 9}, 1009--1021 (1992). 

\bibitem{Ballestetal08ClQGr}  A. Ballesteros,  A. Enciso,    F.J.  Herranz,  and  O. Ragnisco,   
{\rm ``Bertrand spacetimes as Kepler/oscillator potentials"}, 
Class. Quantum Grav. {\bf  25}, no. 16, 165005  (2008).    
 
 \bibitem{RagRig10}   O. Ragnisco  and  O. Riglioni,  
{\rm ``A family of exactly solvable radial quantum systems on space of non-constant curvature with accidental degeneracy in the spectrum"},   
 SIGMA Symmetry Integrability Geom. Methods Appl. {\bf 6},   097  (2010).    

\bibitem{ComRunge}   A. Ballesteros,  A. Enciso,    F.J.  Herranz,  and  O. Ragnisco,   
{\rm ``Hamiltonian systems admitting a Runge-Lenz vector and an optimal extension of Bertrand's theorem to curved manifolds"},    
 Comm. Math. Phys.  {\bf  290},  no. 3, 1033--1049  (2009). 

\bibitem{Evans90}   N.W.  Evans,  
{\rm ``Super-integrability of the Winternitz system"},  
Phys. Lett. A {\bf 147}, no. 8--9,  483--486 (1990).

\bibitem{Evans91}   N.W.  Evans,  
{\rm ``Group theory of the Smorodinsky-Winternitz system"}, 
J. Math. Phys. {\bf 32}, no. 12,  3369--3375 (1991).

\bibitem{LetterBH}  A. Ballesteros   and  F.J.  Herranz,  
{\rm ``Universal integrals for superintegrable systems on N-dimensional spaces of constant curvature"},  
J. Phys. A: Math. Theor. {\bf 40}, no. 2, F51--F59  (2007).

 \bibitem{ChanuDgR11}  C. Chanu,  L.  Degiovanni,  and G. Rastelli,  
{\rm ``First integrals of extended Hamiltonians in n+1 dimensions generated by powers of an operator"},  
SIGMA Symmetry Integrability Geom. Methods Appl.  {\bf  7}, Paper 038,  (2011). 

\bibitem{ChanuDgR14}  C. Chanu,  L.  Degiovanni,  and G. Rastelli,    
{\rm ``Extensions of Hamiltonian systems dependent on a rational parameter"}, 
J. Math. Phys.  {\bf 55},  no. 12,  122703  (2014).

 \bibitem{ChanuDgR15} C. Chanu,   L.  Degiovanni,  and G. Rastelli,  
{\rm ``Warped product of Hamiltonians and extensions of Hamiltonian systems"}, 
 J. Phys. Conf. Ser.  {\bf  597},  012024  (2015).

\bibitem{Thom86}  G.  Thompson, 
 {\rm ``Killing tensors in spaces of constant curvature"},  
 J. Math. Phys. {\bf  27}, no. 11, 2693--2699  (1986).  

\bibitem{BenChaRast01}  S.  Benenti,  C.  Chanu,  and  G.  Rastelli,  
 {\rm ``Variable separation for natural Hamiltonians with scalar  and vector potentials on Riemannian manifolds"},   
J. Math. Phys. {\bf  42},  no. 5, 2065--2091  (2001). 

\bibitem{ChadGMc06} C.  Chanu,  L.  Degiovanni,  and  R.G. McLenaghan, 
{\rm ``Geometrical classification of Killing tensors on bidimensional flat manifolds"}, 
J. Math. Phys.  {\bf 47}, 073506 (2006).  

\bibitem{HorMcLSm09}   J.T.  Horwood,  R.G. McLenaghan, and R.G.  Smirnov, 
{\rm ``Hamilton-Jacobi theory in three-dimensional Minkowski space via Cartan geometry"},   
J. Math. Phys. {\bf  50},  053507  (2009). 

\bibitem{RajaratMcLen14}   K. Rajaratnam  and   R.G.   McLenaghan, 
{\rm ``Killing tensors, warped products and the orthogonal separation of the Hamilton-Jacobi equation"}, 
 J. Math. Phys. {\bf  55},  no. 1,  013505  (2014). 

 \bibitem{Grav04Jmp} S.  Gravel,  
 {\rm ``Hamiltonians separable in Cartesian coordinates and third-order integrals of motion"},  
 J. Math. Phys. \textbf{ 45},  no. 3, 1003-1019  (2004).      

\bibitem{MW08JPa}  I. Marquette  and  P. Winternitz, 
 {\rm ``Superintegrable systems with third-order integrals of motion"}, 
 J. Phys. A: Math. Theor.   \textbf{ 41},  no. 30,  304031  (2008).  

\bibitem{TW10JPa}  F. Tremblay  and  P.  Winternitz, 
 {\rm ``Third-order superintegrable systems separating in polar coordinates"},   
 J. Phys. A  Math. Theor. \textbf{ 43},  no. 17,  175206  (2010).  

\bibitem{MatShe11} V.S. Matveev and  V.V. Shevchishin,
``Two-dimensional superintegrable metrics with one linear and one cubic integral", 
J. Geom. Phys.  {\bf  61},  no. 8,  1353-1377  (2011).      

\bibitem{CCR13JPa}   R. Campoamor-Stursberg,  J.F.  Cari\~nena,  and  M.F.  Ra\~nada, 
``Higher-order superintegrability of a Holt related potential", 
 J. Phys. A: Math. Theor.  {\bf  46},  no. 43,  435202  (2013).      

\bibitem{vHolt07}   J.W.~van Holten,   
``Covariant Hamiltonian dynamics",
Phys. Rev. D  {\bf 75}, 025027 (2007).   

\bibitem{Visien10}  M. Visinescu,  
 ``Higher order first integrals of motion in a gauge covariant Hamiltonian framework", 
 Mod. Phys.  Lett. A  {\bf 25},  341-350 (2010).   

\bibitem{Redm64}   P.J.  Redmond,  
{\rm ``Generalisation of the Runge-Lenz vector in the presence of an electric field"},   
 Phys. Rev. (2) {\bf  133},  B1352--B1353  (1964). 

\bibitem{LeachGor88}  P.G.L.  Leach  and  V.M.  Gorringe, 
{\rm ``A conserved Laplace-Runge-Lenz-like vector for a class of three-dimensional motions"}, 
Phys. Lett. A  {\bf  133}, no. 6, 289--294 (1988).

\bibitem{HolasMar90}  A. Holas   and     N.H.  March,   
{\rm ``A generalisation of the Runge-Lenz constant of classical motion in a central potential"},    
 J. Phys. A : Math. Gen. {\bf  23}, no. 5, 735--749  (1990). 

\bibitem{LeachFle03}  P.G.L.  Leach   and    G.P. Flessas, 
{\rm ``Generalisations of the Laplace-Runge-Lenz vector"}, 
 J. Nonlinear Math. Phys. {\bf  10},   no. 3,  340--423.  (2003).
 
\bibitem{CRS08}  J.F.  Cari\~nena,  M.F.  Ra\~nada, and M. Santander,    
{\rm ``The Kepler problem and the Laplace-Runge-Lenz vector on spaces of constant curvature and arbitrary signature"},  
 Qual. Theory Dyn. Syst.  {\bf   7}, no. 1, 87--99  (2008). 

\bibitem{White10}  H. White,   
{\rm ``On a class of dynamical systems admitting both Poincar\'e and Laplace-Runge-Lenz vectors"},    
 Nuovo Cimento B  {\bf  125}, no. 1, 7--25 (2010). 
 
\bibitem{BenYaacov10}  U.  Ben-Yaacov,
{\rm ``Laplace-Runge-Lenz symmetry in general rotationally symmetric systems"},  
 J. Math. Phys. {\bf 51},  122902 (2010).  

\bibitem{MarquetteJPA10}   I.  Marquette,  
{\rm ``Superintegrability and higher order polynomial algebras"},   
 J. Phys. A: Math. Gen.  {\bf 43}, no. 13,  135203 (2010). 
 
\bibitem{MarquetteJMP10}   I.  Marquette,  
{\rm ``Generalised MICZ-Kepler system, duality, polynomial, and deformed oscillator algebras"},    
 J. Math. Phys. {\bf 51}, 102105 (2010). 
 
\bibitem{Sigma11} A.  Ballesteros,   A. Enciso,    F.J.  Herranz,   O. Ragnisco,  and  O. Riglioni, 
{\rm ``Superintegrable oscillator and Kepler systems on spaces of nonconstant curvature via the S\"ackel Transform"}, 
 SIGMA  Symmetry Integrability Geom. Methods Appl. {\bf  7},  048  (2011).

\bibitem{Nikitin14}   A.G.  Nikitin, 
{\rm ``Laplace-Runge-Lenz vector with spin in any dimension"}, 
J. Phys. A: Math. Theor.  {\bf  47},  no. 37,  375201  (2014).
 
\bibitem{Ra15Jmp}  M.F.  Ra\~nada,   
{\rm ``Superintegrable deformations of superintegrable systems : Quadratic superintegrability and higher-order superintegrability"},
J. Math. Phys. {\bf 56}, no. 4,  042703 (2015). 

\bibitem{PhysD2008}   A.  Ballesteros,  A. Enciso,   F.J.  Herranz,  and  O. Ragnisco,
{\rm ``A maximally superintegrable system on an n-dimensional space of nonconstant curvature"},
Physica D {\bf 237}, no. 4,  505--509  (2008).
 
\bibitem{Bal14AnnPhys}   A.  Ballesteros,   A. Enciso,    F.J.  Herranz,  O.  Ragnisco,  and O.  Riglioni, 
{\rm ``An exactly solvable deformation of the Coulomb problem associated with the Taub--NUT metric"},
Ann. Phys.   {\bf 351},  540--557  (2014).
  
\bibitem{IwaiKat94}   T.  Iwai and  N. Katayama, 
{\rm ``Two kinds of generalised Taub-NUT metrics and the symmetry of associated dynamical systems"},  
J. Phys. A: Math. Gen. {\bf 27}, no. 9, 3179--3190 (1994).

\bibitem{CGMS14}  J.F.  Cari\~{n}ena,  I.  Gheorghiu,  E. Mart\'{\i}nez,  and  P. Santos,
{\rm ``Conformal Killing vector fields and a  virial theorem"}, 
 J. Phys. A: Math. Theor. {\bf 47},  465206   (2014). 
  
 \bibitem{CaLop91} J.F.  Cari\~{n}ena   and C.  L\'opez,  
{\rm ``Symplectic Structure on the set of geodesics of a Riemannian manifold"}, 
Int. J. Mod. Phys. A, {\bf 6}, 431--444 (1991). 

 \bibitem{Ben97}  S.  Benenti,  
{\rm ``Intrinsic characterization of the variable separation in the Hamilton--Jacobi equation"},
J. Math. Phys. {\bf  38}, 6578--6602 (1997). 

 


\end{thebibliography}
\end{document}